\documentclass[a4paper,11pt]{article}
\usepackage{stmaryrd}
\usepackage{amsfonts}
\usepackage{bbm}
\usepackage{float} 
\usepackage{float}
\usepackage{amscd}
\usepackage{mathrsfs}
\usepackage{latexsym,amssymb,amsmath,amscd,amscd,amsthm,amsxtra,xypic}
\usepackage{graphicx}
\usepackage[utf8]{inputenc}
\usepackage[T1]{fontenc}
\usepackage{lmodern}
\usepackage{amssymb}
\usepackage[all]{xy}
\usepackage{nicefrac,mathtools,enumitem}
\usepackage{microtype}
\usepackage{lipsum}
\usepackage{mathtools}
\usepackage{xfrac}

\usepackage{geometry}
\usepackage{hyperref}

\usepackage{times}
\usepackage{amsthm}
\usepackage{amsmath}
\usepackage{amsfonts}
\usepackage{amssymb}
\usepackage{mathdots}
\usepackage{color}
\usepackage{mathrsfs}
\usepackage{stmaryrd}
\SetSymbolFont{stmry}{bold}{U}{stmry}{m}{n}

\usepackage{bm}
\usepackage{tikz}
\usetikzlibrary{calc,math}

\usepackage{tikz-cd}
\usepackage{tikz}
\usetikzlibrary{matrix,arrows,decorations.pathmorphing}
\usepackage[all]{xy}
\usepackage{graphicx}
\usepackage{subfigure}

\usepackage[toc,page]{appendix}
\usepackage[usestackEOL]{stackengine}

\usepackage{theoremref}
\newtheorem{thm}{Theorem}[section]
\newtheorem{lem}[thm]{Lemma}

\newtheorem{pro}[thm]{Proposition}

\theoremstyle{definition}

\newtheorem{rmk}[thm]{Remark}
\newtheorem{defi}[thm]{Definition}
\newtheorem*{assumption*}{Assumption}

\newcommand{\be }{\begin{equation}}
\newcommand{\ee }{\end{equation}}

\newcommand{\norm}[1]{\left\Vert#1\right\Vert}

\newcommand{\h}{\mathfrak h}

\def\qed{\hfill ~\vrule height6pt width6pt depth0pt}

\newcommand{\br}[1]{   [ \cdot,    \cdot  ]   }

\newcommand {\Sect}{{\rm Sect}}

\newcommand{\diag}{\mathrm{diag}}

\newcommand{\I}{{\mathrm{i}}}
\usepackage{pgfplots}
\pgfplotsset{compat=1.18}
\title{On a factorization of Stokes matrices at
a third order pole via isomonodromy deformation}
\author{Xiaomeng Xu}
\date{}

\newcommand{\Addresses}{{
  \bigskip
  \footnotesize
\noindent \textsc{
School of Mathematical Sciences \& Beijing International Center
for Mathematical Research, Peking University, Beijing 100871, China}\par\nopagebreak
  \textit{E-mail address}: \texttt{xxu@bicmr.pku.edu.cn}
}}
\usepackage{booktabs}
\newcommand{\off}{\mathrm{off}}

\newcommand{\out}{\mathrm{out}}

\newcommand{\T}{\mathcal T}
\allowdisplaybreaks[1]
\hypersetup{hidelinks}

\begin{document}
\maketitle
\begin{abstract}
We study the Stokes matrices of meromorphic $n\times n$ linear systems of ordinary differential equations with
a third order pole and skew-Hermitian coefficients.
Along a one parameter isomonodromy deformation, we prove that
the coefficient matrices admit regularized limits as the deformation
parameter tends to infinity. In this limit, the original system
degenerates on different scales into a system with a pole of order
two and a family of explicitly solvable $2\times 2$ systems. Our main result expresses the Stokes matrices of the original system in terms of those of the limiting systems.
\end{abstract}

\section{Introduction}\label{sec:introduction}

In this paper, we study the Stokes matrices of the meromorphic
$n\times n$ linear system
\begin{equation}\label{eq1}
\frac{dF}{dz}=\left(Uz+A+\frac{B}z\right)F, \qquad A=D+N,
\end{equation}
where 
\begin{equation}\label{UDdefine}
 U=\sqrt{-1}\cdot \diag(u_1,\ldots,u_n),\quad
D=\sqrt{-1}\cdot\diag(d_1,\ldots,d_n),\qquad
 u_1<\cdots<u_n,\quad d_1<\cdots<d_n
\end{equation}
are diagonal matrices with purely imaginary eigenvalues, $A,B\in\mathfrak u(n)$ are $n \times n$ skew-Hermitian matrices, and $D$ and $N$ are the diagonal and off-diagonal parts of $A$ respectively. It has a third order pole at $\infty$ and a simple pole at $z=0$. Its irregular type at $\infty$ is $Uz^2/2+Dz$. In Section \ref{arborder} we discuss an extension to higher order poles. 

Standard resummation theory (see e.g., \cite{Balser, LR, Wa}) gives sectorial regions around $z=\infty$, on each of which there is a unique (therefore canonical) holomorphic solution with the prescribed formal asymptotic expansion. The transition matrices between these solutions are encoded by four Stokes matrices, see Section \ref{Stokesmatrices},
\[S_{0}(U,D;N,B),\ldots, S_{3}(U,D;N,B)\in {\rm GL}(\mathbb{C}^n).\]
The Stokes matrices are highly transcendental functions of the parameters $U,D$ and $N,B$. One way to understand them is to obtain their
explicit expansions near singular points of the space of irregular types $U,D$. In the paper we replace $D$ by $tD$ with $t\in\mathbb{R}$, and let the coefficient matrices $N(t), B(t)$ vary according to isomonodromy deformation
equation in $t$ (with $N(1)=N, B(1)=B$). Along this deformation, the Stokes matrices $S_{k}(U,tD;N(t),B(t))$ are independent of $t$.
In Theorem \ref{thm1}, we prove that as $t\rightarrow+\infty$, $N(t), B(t)$ admit regularized limits $N_0,B_0$, called boundary values. We then show that as $t\rightarrow+\infty$, the
linear system decouples into an $n\times n$ linear system with second order pole and multiple explicitly solvable $2\times 2$ systems with boundary values as coefficient matrices. In Theorem \ref{thm2}, we give a simple factorization of the original Stokes matrices via those of the limiting systems. 

For $n=2$, the isomonodromy equation and the linear system reduce to the Painlev\'e IV equation and its Lax pair. Isomonodromy methods for asymptotic and connection problems
have played an important role in the theory of Painlev\'e equations, for example,
see Jimbo \cite{Jimbo1982}, the books by Its and Novokshenov \cite{IN}, Fokas, Its, Kapaev and Novokshenov \cite{FIKN} and the references therein. In our recent works, the approach has been applied to the study of Stokes matrices of meromorphic linear systems with one or two second order poles \cite{Xu, TXu, WangXu}, and of linear systems with noncommutative coefficients and arbitrary order poles \cite{Xu2026}. In this paper, we apply the approach to third order pole case. Our motivation for taking $U,A,B\in\mathfrak{u}(n)$ comes from the WKB approximation of irregular Knizhnik–Zamolodchikov equations \cite{Re, FR, Xu2026}, whose coefficient matrices are skew-Hermitian. For  geometric and representation theoretic interpretations of
this skew-Hermitian condition, see \cite{ANXZ, Xu, Xu20264}.

\subsection{Isomonodromy equations and boundary values}
We fix $U$ and $D$, and consider the one parameter isomonodromy
deformation with irregular parameters $(U,tD)$, $t\in\mathbb R$,
along which the Stokes matrices
$S_k(U,tD;N(t),B(t))$ remain constant. 
The following corresponding compatible linear system is a specialization
of the equations of Jimbo-Miwa-Ueno 
\cite[Example 3.3]{JMU}. 

We consider the linear system for a matrix function $F(z,t)$ 
\begin{align}\label{zeq}
\frac{\partial F}{\partial z}&=\left(Uz+A(t)+\frac{B(t)}z\right)F,
 \qquad A(t)=tD+N(t),\\ \label{teq}
\frac{\partial F}{\partial t}&=(Dz+\T N(t))F.
\end{align}
Here ${\rm diag}(N(t))=0$, and $\T N\in \mathfrak u(n)$ denotes the matrix whose entries are
\begin{equation}
    (\T N)_{ij}:=
\left\{\begin{array}{ll}
    \frac{d_i-d_j}{u_i-u_j}N_{ij}, &  i \neq j\\
    0, &  i = j
  \end{array}\right..
\end{equation}
The compatibility condition gives the nonlinear ordinary differential equations in $t$,
\begin{align}\label{isoNeq}
\frac{d N}{d t}&=-t[D,\T N]+[D,B]+[\T N,N],\\ \label{isoBeq}
\frac{d B}{d t}&=[\T N,B].
\end{align}

Our first result gives the regularized limits of solutions of
\eqref{isoNeq}--\eqref{isoBeq} as $t\to+\infty$ along the real axis. We make a generic assumption on the diagonal matrices $U$ and $D$ that
\begin{equation}\label{separation}
 \frac{d_i-d_j}{u_i-u_j}\ne\frac{d_k-d_l}{u_k-u_l}
 \quad\text{for }i<j,\ k<l,\ (i,j)\ne(k,l).
\end{equation}
Let $\mathfrak u(n)$ denote the space of skew-Hermitian matrices and let
$\mathfrak u(n)_{\off}$ denote the subspace of off-diagonal ones.
\begin{thm}\label{thm1}
Assume \eqref{UDdefine} and \eqref{separation}. For every
$(N_*,B_*)\in\mathfrak u(n)_{\off}\times\mathfrak u(n)$, the initial
value problem for the isomonodromy equation \eqref{isoNeq}--\eqref{isoBeq} with
$(N(1),B(1))=(N_*,B_*)$ has a unique global real analytic solution
\[
 (N,B):t\in\mathbb R\longrightarrow
       \mathfrak u(n)_{\off}\times\mathfrak u(n).
\]
This solution has a unique boundary value $(N_0,B_0)\in\mathfrak u(n)_{\off}\times\mathfrak u(n)$, characterized by the following asymptotics as $t\to+\infty$,
\begin{align}\label{Nlimit}
 e^{\frac{\I(d_i-d_j)^2t^2}{2(u_i-u_j)}}
 t^{-\big(\Lambda_i-\Lambda_j
    -\frac{2(N_0)_{ij}(N_0)_{ji}}{\I(u_i-u_j)}\big)}
 N_{ij}(t)
 &=(N_0)_{ij}+O(t^{-1}),\qquad i\ne j,\\ 
 \label{Blimit}
 t^{2\diag (B_0)-\Lambda}\cdot B(t)\cdot t^{-(2\diag (B_0)-\Lambda)}
 &=B_0+O(t^{-1}).
\end{align}
Here $t^X={\rm exp}(X{\rm log}(t))$ for $t>0$ with ${\rm log}(t)\in\mathbb{R}$, and $\Lambda={\rm diag}(\Lambda_1,\ldots,\Lambda_n)$ with $\Lambda_i=(B_0)_{ii}+\sum_{j\ne i}\frac{(N_0)_{ij}(N_0)_{ji}}{\I(u_i-u_j)}.$
\end{thm}
Theorem \ref{thm1} is proved in
Section \ref{sec:boundary}. It assigns a boundary value, or asymptotic parameters at $t=+\infty$, to
each skew-Hermitian matrix valued solution of the isomonodromy equation.

\begin{rmk}
For fixed $U,D$ satisfying \eqref{UDdefine} and \eqref{separation},
one can also prove that every pair
$(N_0,B_0)\in\mathfrak u(n)_{\off}\times\mathfrak u(n)$
is the boundary value of a unique real analytic solution
of \eqref{isoNeq}--\eqref{isoBeq}, and that the map
\[
 (N(1),B(1))\longmapsto(N_0,B_0)
\]
is a real analytic diffeomorphism of
$\mathfrak u(n)_{\off}\times\mathfrak u(n)$.
It can be proved directly by solving the asymptotic boundary
problem.
Theorem \ref{thm2} also provides a Riemann-Hilbert
approach, as in \cite{TXu}.
\end{rmk}

\subsection{Stokes matrices and their factorization at $t=+\infty$}
Let us fix a real analytic solution $(N(t),B(t))\in \mathfrak u(n)_{\off}\times\mathfrak u(n)$ with boundary value $(N_0,B_0)$, and consider the linear system \eqref{zeq}. 
The regularized limits in
Theorem \ref{thm1} give its limiting equations in two kinds
of scales (see Sections \ref{secztdeg} and \ref{secijdeg} for more details): near $z=0$, set
$\zeta=tz$, after the diagonal normalization used in
\eqref{Blimit}, the equation \eqref{zeq} converges to
the $n \times n$ linear system
\begin{equation}\label{introoutereq}
 \frac{d Y}{d\zeta}
 =\left(D+\frac{B_0}{\zeta}\right)\cdot Y.
\end{equation}
For any pair $i<j$, near each crossing point
$z=-t(d_i-d_j)/(u_i-u_j)$, where the $i$-th and $j$-th
diagonal entries of $Uz+tD$ coincide, set $
 \eta=z+t\frac{d_i-d_j}{u_i-u_j}.$
After the normalization determined by \eqref{Nlimit}
and a suitable transform, the equation \eqref{zeq}
converges to 
the $n\times n$ linear system
\begin{equation}\label{intronijeq}
 \frac{dK^{(ij)}}{d\eta}
 =\left(U\eta+(N_0)_{ij}E_{ij}+(N_0)_{ji}E_{ji}\right)\cdot K^{(ij)},
\end{equation}
where $E_{ab}$ is the matrix unit with its only nonzero entry,
equal to one, in position $(a,b)$. It is an exactly solvable system, see Section \ref{cat}.

We denote the Stokes matrices of \eqref{zeq} and \eqref{intronijeq} by $S_{k}(U,tD;N(t),B(t))$ and $S_k^{(ij)}(U,N_0)$, $k=0,1,2,3$ respectively, and denote by $S_{\out}(D,B_0)\in {\rm GL}_n(\mathbb{C})$ the Stokes matrix of \eqref{introoutereq} used below. Following \cite{JMU}, the matrices $S_{k}(U,tD;N(t),B(t))$ are independent of $t\in\mathbb{R}$. Moreover, the skew-Hermitian condition determines $S_2, S_3$ from $S_0,S_1$. See Section \ref{Stokesmatrices} for more details.

Our second result gives a factorization of the Stokes matrices of the linear system \eqref{zeq} into the Stokes matrices of the limiting systems. 
Order the pairs $(p_a,q_a)$, $p_a<q_a$, such that
\begin{equation}\label{orderpq}
 \frac{d_{p_1}-d_{q_1}}{u_{p_1}-u_{q_1}}>
 \cdots>
 \frac{d_{p_m}-d_{q_m}}{u_{p_m}-u_{q_m}},
 \qquad m=\frac{n(n-1)}2.
\end{equation}
Then we have
\begin{thm}\label{thm2}
The Stokes
matrices of \eqref{zeq} are expressed in terms of those of
\eqref{introoutereq} and \eqref{intronijeq} by
\begin{equation}\label{factS}
S_0S_1=D_{\out}^{-1}\cdot S_{\out}(D,B_0)\cdot D_{\out}
 \cdot T^{(p_mq_m)}\cdots T^{(p_1q_1)},
\end{equation}
where $T^{(ij)}$ is defined by 
\begin{equation}\label{Tmatrix}
 T^{(ij)}:=(D^{(ij)})^{-1}\cdot
 S_0^{(ij)}S_1^{(ij)}\cdot D^{(ij)},
\end{equation}
and $D_{\out},D^{(ij)}$ are $n\times n$ diagonal matrices with entries
\begin{equation}\label{Dmatrix}
 \begin{aligned}
 (D_{\out})_{kk}
 &=\prod_{l\ne k}
   \left(\frac{d_k-d_l}{u_k-u_l}\right)^{\frac{(N_0)_{kl}(N_0)_{lk}}{\I(u_k-u_l)}},\\[4pt]
 (D^{(ij)})_{kk}&=
 \Big(-\frac{d_i-d_j}{u_i-u_j}\Big)^{(B_0)_{kk}}
 \prod_{\substack{l\ne k\\\{k,l\}\ne\{i,j\}}}
 \Big(\frac{d_k-d_l}{u_k-u_l}-\frac{d_i-d_j}{u_i-u_j}\Big)^{\frac{(N_0)_{kl}(N_0)_{lk}}{\I(u_k-u_l)}},
 \quad 1\le k\le n.
 \end{aligned}
\end{equation}
The product omits $l=j$ when $k=i$ and $l=i$ when $k=j$.
\end{thm}
The matrices $S_0,S_1$ are the unique lower and upper
unitriangular factors of the right hand side of
\eqref{factS}, and
$S_2,S_3$ are then determined by \eqref{Sconj}. In the paper, we use the convention that
$\log a$ is real for $a>0$, and
$\log a=\log|a|+\pi\I$ for $a<0$.

\subsection{Explicit expression of the Stokes matrices as $t\rightarrow+\infty$ and $d_1\ll\ldots\ll d_n$}\label{cat}
Theorem \ref{thm2} states that if $N(t), B(t)$ are solutions of \eqref{isoNeq}-\eqref{isoBeq} with boundary values $N_0,B_0$ at $t=+\infty$, then the Stokes matrices
$S_k(U,tD;N(t),B(t))$ of \eqref{zeq}, independent of $t$, are given by an explicit expression of the Stokes matrices $S_{\out}(D,B_0)$ and $S_k^{(ij)}(u_i,u_j;(N_0)_{ij},(N_0)_{ji})$ of the equations \eqref{introoutereq} and \eqref{intronijeq} for all $1\le i<j\le n$.

The equation \eqref{intronijeq} is exactly solvable. It decouples into $n-2$ scalar equations and the $2 \times 2$ system
\begin{equation}\label{ijeq}
 \frac{d \mathcal K}{d\eta}
 =\left(\begin{pmatrix}
 \I u_i&0\\
 0&\I u_j
 \end{pmatrix}\eta+\begin{pmatrix}
 0&(N_0)_{ij}\\
 (N_0)_{ji}&0
 \end{pmatrix}\right)\cdot \mathcal K.
\end{equation}
The $2\times 2$ system reduces to a parabolic cylinder equation. The connection formula for parabolic cylinder functions, see e.g., \cite{OLBC}, gives the Stokes matrices.
In particular, two $n\times n$ Stokes matrices of \eqref{intronijeq} are
\[
 S_0^{(ij)}={\rm Id}_n+(S_0^{(ij)})_{ji}E_{ji},
 \qquad
 S_1^{(ij)}={\rm Id}_n+(S_1^{(ij)})_{ij}E_{ij},
\]
where
 \begin{align}\label{Sentry}
 (S_0^{(ij)})_{ji}
 &=-\frac{\I\sqrt{2\pi}\,(N_0)_{ji}}
 {\Gamma\!\left(1-\frac{(N_0)_{ij}(N_0)_{ji}}{\I(u_i-u_j)}\right)}
 \bigl(\I(u_i-u_j)\bigr)^{
 -\frac{(N_0)_{ij}(N_0)_{ji}}{\I(u_i-u_j)}-\frac12},\\
 (S_1^{(ij)})_{ij}
 &=e^{-\frac{\pi(N_0)_{ij}(N_0)_{ji}}{u_i-u_j}}\frac{\sqrt{2\pi}\,(N_0)_{ij}}
{\Gamma\!\left(1+\frac{(N_0)_{ij}(N_0)_{ji}}{\I(u_i-u_j)}\right)}
 \bigl(\I(u_i-u_j)\bigr)^{
 \frac{(N_0)_{ij}(N_0)_{ji}}{\I(u_i-u_j)}-\frac12}.
 \end{align}
Here for $i<j$, we have $u_i<u_j$ and $\log(\I(u_i-u_j))=\log(u_j-u_i)-\frac{\pi\I}{2}.$

Therefore, Theorem \ref{thm2} reduces the study of the Stokes matrices to the system \eqref{Yeq} with a second order pole. For fixed $U,D$ let $(N_0,B_0)$ be the boundary value at $t=+\infty$. We can then consider the isomonodromy equation of \eqref{Yeq} with respect to $D$, and take its boundary value $B_{0,cat}\in\frak u(n)$ at $d_1\ll \ldots \ll d_n$ as in \cite{Xu}. Following \cite{Xu}, $S_{out}(D,B_0)$ is expressed explicitly via $B_{0,cat}$. Inserting this expression and \eqref{Sentry} into \eqref{factS} gives an explicit formula of $S_k(U,D;N(U,D), B(U,D))$ in terms of $N_0,B_{0,cat}$. 

\subsection{Poles of arbitrary order}\label{arborder}
The same construction suggests a recursive approach to the study of Stokes matrices at poles of
order $m+2$. Consider the $n\times n$ system
\begin{equation}\label{meq}
 \frac{d F}{dz}
 =\left(
 Uz^m+(tD+A(t))z^{m-1}
 +\sum_{a=0}^{m-2}C_a(t)z^a+\frac{B(t)}z
 \right)F,
\end{equation}
where $U=\I\diag(u_1,\ldots,u_n)$ and
$D=\I\diag(d_1,\ldots,d_n)$ satisfy suitable generic conditions.
The coefficients $A(t), C_a(t), B(t)$ are skew-Hermitian, $\diag (A)=0$, and evolve
according to the isomonodromy equation \cite{JMU} in one parameter $t\in \mathbb{R}$.

As in the case $m=1$ treated in this paper, the leading diagonal
term $Uz^m+tDz^{m-1}$ suggests two kinds of scales.
On the first scale $z=t^{-1/m}\zeta$, near $z=0$, 
after normalization 
we expect a limiting equation of the form
\begin{equation}\label{m+1pole}
 \frac{dY}{d\zeta}
 =\left(
 D\zeta^{m-1}+\sum_{a=0}^{m-2}Q_{a,0}\zeta^a
 +\frac{B_0}{\zeta}
 \right)Y,
\end{equation}
which has a pole of order $m+1$ at infinity. The coefficients
$Q_{a,0}$ and $B_0$ should arise from suitable regularized limits of
$C_a(t)$ and $B(t)$. On the second scale
 $z=z_{ij}(t)+t^{-(m-1)/2}\eta,$
where $z_{ij}(t)$ is the simple zero near $-t(d_i-d_j)/(u_i-u_j)$ of the difference of the $i$-th and $j$-th
entries of the diagonal norm form. After
proper normalization we expect to get a limiting equation that decouples into $n-2$ scalar equations and a $2\times 2$ block 
\[
 \frac{dK^{(ij)}}{d\eta}
 =\left(
 \frac{\I(u_i-u_j)}2
 \Big(-\frac{d_i-d_j}{u_i-u_j}\Big)^{m-1}
 \begin{pmatrix}1&0\\0&-1\end{pmatrix}\eta
 +\begin{pmatrix}
 0&(L_0)_{ij}\\
 (L_0)_{ji}&0
 \end{pmatrix}
 \right)K^{(ij)}.
\]

Then like the $m=1$ case, comparing canonical solutions on these scales should relate the
monodromy data of \eqref{meq} to those of the system \eqref{m+1pole} and
the $2\times 2$ exactly solvable systems. An iteration would give an explicit formula of Stokes matrices in terms of boundary values. 

\subsection{Applications: WKB approximation, spectral curves, crystal basis and Riemann-Hilbert-Birkhoff maps}
For a pole of order two, expressions for the Stokes matrices in terms of boundary values have been used to study the relation of WKB approximation to spectral curves and crystal basis \cite{ANXZ,Xu,Xu20264}. The result in the present paper suggests analogous applications for higher order poles.

\paragraph{Spectral curves and WKB approximation.} The factorization of the equation and the Stokes matrices, studied in this paper, has an interpretation
in terms of its spectral curve. Consider the spectral curve determined by the characteristic polynomial
\[
 \Gamma_t:\quad
 \det\left(\lambda {\rm Id}_n-Uz-tD-N(t)-\frac{B(t)}z\right)=0.
\]
The regularized limits in Theorem \ref{thm1} determine its
local geometry on the two scales.
Near $z=0$, with $\zeta=tz$ and $\nu=\lambda/t$, the characteristic equation
converges to 
the one of the limiting system \eqref{introoutereq},
\[
 \Gamma_{\out}:\quad
 \det\left(\nu {\rm Id}_n-D-\frac{B_0}{\zeta}\right)=0.
\]
Near the crossing point associated with $i<j$, with $\eta=z+t\frac{d_i-d_j}{u_i-u_j}$ and $\xi=\lambda-\I t\,\frac{u_i d_j-u_j d_i}{u_i-u_j}$, the two corresponding branches have the limiting equation
\[
 \Gamma_{ij}:\quad
 \det\left(\xi {\rm Id}_2-\begin{pmatrix}
 \I u_i& 0\\
 0&\I u_j
\end{pmatrix}\eta-\begin{pmatrix}
 0 &(N_0)_{ij}\\
 (N_0)_{ji}&0
 \end{pmatrix}\right)=0.
\]
On the scale $x=z/t,\ y=\lambda/t$, the limiting equation for fixed $x\ne 0$ is $\prod_{i=1}^n (y-\I u_ix-\I d_i)=0$ with rational components $y=\I u_ix+\I d_i$ denoted by $R_1,\ldots,R_n$. Then
for a generic boundary value $(N_0,B_0)$, the limits
suggest that, as $t\to+\infty$, the spectral curve $\Gamma_t$
degenerates to a nodal curve consisting of
$\Gamma_{\mathrm{out}}$, rational components
$R_1,\ldots,R_n$, and $\Gamma_{ij}$ for all $i<j$, where
$R_i$ meets $\Gamma_{\mathrm{out}}$ at one node,
while $\Gamma_{ij}$ meets $R_i$ at one node and $R_j$ at another. In view of \cite{ANXZ}, we expect that the periods
over the vanishing cycles associated with $\Gamma_{ij}$ govern the WKB asymptotics of the corresponding
two dimensional factors in \eqref{factS}.

\paragraph{WKB approximation of Stokes matrices of irregular
Knizhnik-Zamolodchikov (KZ) equations.}
A further motivation is the WKB approximation of irregular KZ equations on singular modules for truncated current Lie algebras
\cite{Xu2026,Xu20262}. The skew-Hermitian condition considered here
is adapted to this problem. It follows from \cite{Xu2026,Xu20262} that Stokes matrices of the irregular KZ equations give rise to actions of a $q$-deformation of truncated current Lie algebras on the singular modules. We expect that 
the WKB approximation of the Stokes matrices, in the iterated limits described in Section \ref{arborder}, gives rise to crystal like operators in Gelfand-Tsetlin type basis in the singular modules, extending the result for second order poles \cite{Xu, Xu20264} where the WKB approximation gives rise to the crystal structures of quantum group $U_q(\frak{gl}_n)$.

\paragraph{Riemann-Hilbert-Birkhoff (RHB) maps and linearization of Poisson brackets.} Following \cite{Boalch}, and restricting to the
skew-Hermitian real form, the moduli space $\frak u(n)_{\off}\times\frak u(n)$ of the linear system \eqref{eq1} and the space $U_-\times U_+\times \h(\I\mathbb{R})$ of Stokes matrices, i.e., the de Rham and Betti moduli spaces, carry natural Poisson
structures, obtained by Hamiltonian
and quasi-Hamiltonian reduction respectively.
For fixed $U, D$ the RHB map
\[\mu(U,D):\frak u(n)_{\off}\times\frak u(n)\rightarrow U_-\times U_+\times \h(\I\mathbb{R})~;~(N,B)\mapsto (S_0,S_1,\Lambda)\]
is a Poisson diffeomorphism, where $U_\pm$ are the spaces of complex upper and lower
unitriangular matrices. Theorems \ref{thm1} and \ref{thm2} express the RHB map via the boundary value,
\[
\mu_0: (N_0,B_0)
 \rightarrow
 (N(1),B(1))
 \xrightarrow{\ \mu(U,D)\ }
 (S_0,S_1,\Lambda),
\]
Since the isomonodromy flow $(N(t), B(t))$ and the normalization in Theorem \ref{thm1} preserve the Poisson structure on $\frak u(n)_{\off}\times\frak u(n)$, taking $t\rightarrow+\infty$ we see that $\mu_0$
is also a Poisson diffeomorphism.
The factorization in Theorem \ref{thm2} expresses $\mu_0$
in terms of the RHB map for the system
\eqref{introoutereq} with a second order pole, together with explicit two dimensional
factors.
The RHB map for \eqref{introoutereq} gives a
Ginzburg-Weinstein linearization of the dual Poisson
Lie group $U(n)^*$ \cite{GW}.
This factorization therefore provides a recursive approach
to linearization problems for the Poisson structures on
Betti moduli spaces.

\section{Boundary values of the solutions of isomonodromy equation}\label{sec:boundary}
Throughout, we use the matrix norm $\|A\|=\sup_{v\in \mathbb{C}^n, \|v\|=1}\|Av\|$. We write $C$ for a positive constant whose
values may vary from one estimate to another. The constants in this section are uniform in $t$ and may depend on $U,D$ and the initial data.

\subsection{Existence and boundedness of real analytic solutions}
The following
lemma uses conserved quantities to give global boundedness for \eqref{isoNeq}-\eqref{isoBeq}. 

\begin{lem}\label{bounds}

For any initial data in $\mathfrak u(n)_{\off}\times\mathfrak u(n)$, the solution $(N(t),B(t))$ of
\eqref{isoNeq}-\eqref{isoBeq} exists and is bounded for all $t\in\mathbb R$. Moreover, the diagonal matrix 
\begin{equation}\label{Lambda}
\Lambda=\diag(\Lambda_1,\ldots,\Lambda_n),\qquad \Lambda_i=B_{ii}(t)+\sum_{j\ne i}\frac{N_{ij}(t)N_{ji}(t)}{\I(u_i-u_j)},
\end{equation}
and the eigenvalues of $B(t)$
are independent of $t$.

\end{lem}
\begin{proof}
The right hand sides of \eqref{isoNeq}-\eqref{isoBeq} are
polynomial in $t,N,B$, so the initial value problem has a unique
local analytic solution. 
Then it follows from $\frac{d_i-d_j}{u_i-u_j}=\frac{d_j-d_i}{u_j-u_i}\in\mathbb R$ that if $N$ is skew-Hermitian, so is $\T N$. Since by \eqref{UDdefine} $U, D\in \mathfrak u(n)$ and the commutator of skew-Hermitian matrices is still skew-Hermitian, the right hand sides of equations \eqref{isoNeq} and \eqref{isoBeq} remain skew-Hermitian. 
Hence the skew-Hermitian initial data
give a skew-Hermitian solution.

Since the diagonal of $[\T N,N]$ vanishes, taking the diagonal part of the equation \eqref{isoNeq}, we see that the off-diagonal condition $\diag (N)=0$
is preserved. A direct differentiation of \eqref{Lambda}
gives $d\Lambda/dt=0$. Also, for $k=1,\ldots,n$,
\[
 \frac{d}{dt}\operatorname{tr}(B^k)
 =k\operatorname{tr}\bigl(B^{k-1}[\T N,B]\bigr)=0.
\]
Hence the eigenvalues of $B$ are constant. In
particular, since $B$ is skew-Hermitian, $\sum_{i,j}|B_{ij}(t)|^2=-\operatorname{tr}(B(t)^2)
=constant$, the norm of $B(t)$ is bounded.

Multiplying the identity for $\Lambda_i$ in \eqref{Lambda} by $u_i$
and summing over $i$, we obtain
\begin{equation*}
 \sum_{i=1}^n u_i(\Lambda_i-B_{ii})
 =\sum_{i<j}\frac{(u_i-u_j)N_{ij}N_{ji}}{\I(u_i-u_j)}
 =\I\sum_{i<j}|N_{ij}|^2,
\end{equation*}
where we use $N_{ji}=-\overline{N_{ij}}$.
Since $\Lambda$ is constant and $B(t)$ is bounded, the identity bounds $N(t)$ uniformly.
Consequently, the solution exists for all real $t$.
\end{proof}

\subsection{Integration by parts and regularized limits}

In order to show that the left hand sides of \eqref{Nlimit} and \eqref{Blimit} converge as $t\rightarrow+\infty$,
we estimate their derivatives. We write the functions in the normalization by
\begin{equation}
q_{ij}=\frac{N_{ij}N_{ji}}{\I(u_i-u_j)} \text{ for } i\ne j,\quad q_{ii}=0.
\end{equation}

\begin{lem}\label{parts}
Let $(N(t),B(t))$ be a solution as in Lemma \ref{bounds}.
For $t\ne 0$, the following identities hold, with $i\ne j$
in \eqref{partialN},
\begin{align}\label{partialN}
\frac{d N_{ij}}{d t}
 +\frac{\I(d_i-d_j)^2}{u_i-u_j}tN_{ij}
 -\frac{\Lambda_i-\Lambda_j-2q_{ij}}{t}N_{ij}
 &=-\left(\frac {d}{dt}+\frac{\I(d_i-d_j)^2}{u_i-u_j}t\right)
 \left(\frac{h_{N,ij}}{t}+\frac{k_{N,ij}}{t^2}\right)
 +\mathcal R_{N,ij},\\ \label{partialB}
\frac{d B}{d t}+\frac{1}{t}[2\diag (B)-\Lambda,B]
 &=-\frac {d}{dt}\left(\frac{h_B}{t}+\frac{k_B}{t^2}\right)
 +\mathcal R_B,
\end{align}
The entries of the matrices $h_N,k_N,h_B,k_B$ are polynomials in entries of
$N$ and $B$, given by $h_{N,ii}=k_{N,ii}=0$, 
\begin{align}\label{hNeq}
h_{N,ij}
 =&-\frac{(u_i-u_j)B_{ij}}{d_i-d_j}
 +\I\sum_{a\ne i,j}
 \frac{N_{ia}N_{aj}}
      {d_a-d_j-\tfrac{d_i-d_j}{u_i-u_j}(u_a-u_j)},
 \qquad i\ne j,\\ \nonumber
 k_{N,ij}
=&-\I\,\frac{u_i-u_j}{d_i-d_j}
\sum_{a\ne i,j}
\frac{B_{ia}N_{aj}+N_{ia}B_{aj}}
     {d_a-d_j-\tfrac{d_i-d_j}{u_i-u_j}(u_a-u_j)}
\\ \label{kNeq}
&-\sum_{a,b \ne i,j}
\frac{N_{ia}N_{ab}N_{bj}}
     {\bigl(d_a-d_j-\tfrac{d_i-d_j}{u_i-u_j}(u_a-u_j)\bigr)
      \bigl(d_b-d_j-\tfrac{d_i-d_j}{u_i-u_j}(u_b-u_j)\bigr)},
\qquad i\ne j,
\end{align}
and
\begin{align*}
 h_{B,ij}={}&\sum_{a\ne i}\frac{N_{ia}B_{aj}}{\I(d_i-d_a)}
             -\sum_{a\ne j}\frac{B_{ia}N_{aj}}{\I(d_a-d_j)},\\
 k_{B,ij}
 =&-\sum_{a,b\ne i}
 \frac{N_{ib}N_{ba}B_{aj}}{(d_i-d_b)(d_i-d_a)}
 -\sum_{a,b\ne j}
 \frac{B_{ia}N_{ab}N_{bj}}{(d_a-d_j)(d_b-d_j)}\\
 &+\sum_{\substack{a\ne i, b\ne j}}
 \frac{N_{ia}B_{ab}N_{bj}}{(d_i-d_a)(d_b-d_j)}
 +B_{ij}\sum_{b\ne j}
 \frac{N_{jb}N_{bj}}{(d_b-d_j)^2}.
\end{align*}
The remainders $\mathcal R_{N,ij}$ and $\mathcal R_{B}$ are polynomial
functions of $(N,B)$ with coefficients of order $t^{-2}$ or $t^{-3}$.
In particular, since $N(t)$ and $B(t)$ are bounded, the remainders
are $O(t^{-2})$ as $t\to+\infty$.
\end{lem}
\begin{proof}
The parameters $U,D$ are fixed, and \eqref{separation} ensures that
every denominator is nonzero. It suffices to compare the coefficients of order $1$ and $t^{-1}$ on both sides of \eqref{partialN} and \eqref{partialB}.

First, differentiating along the solution $(N(t),B(t))$ gives
\begin{align*}
\frac{d N_{ij}}{dt}=&-\frac{\I(d_i-d_j)^2}{u_i-u_j}tN_{ij}
       +\I(d_i-d_j)B_{ij}+\sum_{a\ne i,j}
 \left(\frac{d_i-d_a}{u_i-u_a}-\frac{d_a-d_j}{u_a-u_j}\right)
 N_{ia}N_{aj},\\
 \frac {d}{dt}\left(\frac{h_{N,ij}}{t}\right)
 =&-\sum_{a\ne b}\frac{\I(d_a-d_b)^2}{u_a-u_b}N_{ab}
                   \frac{\partial h_{N,ij}}{\partial N_{ab}}\\
                   &+\frac{1}{t}\left(
 \sum_{a\ne b}\frac{\partial h_{N,ij}}{\partial N_{ab}}
                ([D,B]+[\T N,N])_{ab}
 +\sum_{a,b}\frac{\partial h_{N,ij}}{\partial B_{ab}}[\T N,B]_{ab}
 \right)-\frac{h_{N,ij}}{t^2},\\
\left(\frac {d}{dt}+\frac{\I(d_i-d_j)^2}{u_i-u_j}t\right)
       \frac{k_{N,ij}}{t^2}
=&\frac{1}{t}\left(
 -\sum_{a\ne b}\frac{\I(d_a-d_b)^2}{u_a-u_b}N_{ab}\frac{\partial k_{N,ij}}{\partial N_{ab}}+\frac{\I(d_i-d_j)^2}{u_i-u_j}k_{N,ij}\right)\\
&+\frac1{t^2}
 \sum_{a\ne b}\frac{\partial k_{N,ij}}{\partial N_{ab}}
                 ([D,B]+[\T N,N])_{ab}+\frac1{t^2}\sum_{a,b}
 \frac{\partial k_{N,ij}}{\partial B_{ab}}[\T N,B]_{ab}
 -\frac{2k_{N,ij}}{t^3}.
\end{align*}

Then, in the $N$-equation \eqref{partialN}, the order one term on the right hand side is
\[
 -\frac{\I(d_i-d_j)^2}{u_i-u_j}h_{N,ij}
 +\sum_{a\ne b}\frac{\I(d_a-d_b)^2}{u_a-u_b}N_{ab}
                  \frac{\partial h_{N,ij}}{\partial N_{ab}}.
\]
Substitution of \eqref{hNeq} gives
\[
 \I(d_i-d_j)B_{ij}
 +\sum_{a\ne i,j}
 \left(\frac{d_i-d_a}{u_i-u_a}-\frac{d_a-d_j}{u_a-u_j}\right)
 N_{ia}N_{aj},
\]
which agrees with the order one term on the left hand side.

At order $t^{-1}$, it remains to verify the following identity
\begin{align*}
&\sum_{a\ne b}\frac{\partial h_{N,ij}}{\partial N_{ab}}
       ([D,B]+[\T N,N])_{ab}
 +\sum_{a,b}\frac{\partial h_{N,ij}}{\partial B_{ab}}
       [\T N,B]_{ab}\\
&\quad-
 \sum_{a\ne b}\frac{\I(d_a-d_b)^2}{u_a-u_b}N_{ab}
       \frac{\partial k_{N,ij}}{\partial N_{ab}}
 +\frac{\I(d_i-d_j)^2}{u_i-u_j}k_{N,ij}
 =(\Lambda_i-\Lambda_j-2q_{ij})N_{ij}.
\end{align*}
Substituting \eqref{hNeq} into the first two sums, the terms
involving $B_{ii},B_{jj}$ and the cubic terms
$N_{ia}N_{ai}N_{ij}$ and $N_{ij}N_{ja}N_{aj}$,
with $a\ne i,j$, give
\[
 (B_{ii}-B_{jj})N_{ij}
 +\sum_{a\ne i,j}(q_{ia}-q_{ja})N_{ij}
 =(\Lambda_i-\Lambda_j-2q_{ij})N_{ij},
\]
where we use $\Lambda_i=B_{ii}+\sum_{a\ne i}q_{ia}$
and $q_{ji}=-q_{ij}$.
The remaining terms are
$B_{ia}N_{aj}$ and $N_{ia}B_{aj}$ with $a\ne i,j$, and
$N_{ia}N_{ab}N_{bj}$ with $i,a,b,j$ distinct.
Substitution of \eqref{kNeq} shows that the terms involving
$k_{N,ij}$ cancel these monomials. 

Similarly, substituting the formulas for $h_B$ and $k_B$
and comparing the terms of orders $1$ and $t^{-1}$
gives \eqref{partialB}. Moreover, one checks that the terms of orders $t^{-2}$ and $t^{-3}$ give the remainders
in \eqref{partialN} and \eqref{partialB},
\begin{align*}
 \mathcal R_{N,ij}&=-\frac{h_{N,ij}}{t^2}-\frac{2k_{N,ij}}{t^3}+\frac1{t^2}\left(
 \sum_{a\ne b}\frac{\partial k_{N,ij}}{\partial N_{ab}}
                  ([D,B]+[\T N,N])_{ab}
 +\sum_{a,b}\frac{\partial k_{N,ij}}{\partial B_{ab}}[\T N,B]_{ab}
 \right),\\
 \mathcal R_{B,ij}&=-\frac{h_{B,ij}}{t^2}-\frac{2k_{B,ij}}{t^3}+\frac1{t^2}\left(
 \sum_{a\ne b}\frac{\partial k_{B,ij}}{\partial N_{ab}}
                  ([D,B]+[\T N,N])_{ab}
 +\sum_{a,b}\frac{\partial k_{B,ij}}{\partial B_{ab}}[\T N,B]_{ab}
 \right).
\end{align*}
This proves the lemma.
\end{proof}

\subsection{Proof of Theorem \ref{thm1}}
By Lemma \ref{bounds}, the solution is global and $N(t),B(t)$ are bounded.
The $h_N,k_N,h_B,k_B$  in Lemma \ref{parts} are therefore bounded, and the
remainders $\mathcal R_{B,ij}, \mathcal R_{N,ij}$ are $O(t^{-2})$.

\paragraph{Limits of $B_{ii}$ and $N_{ij}N_{ji}$.} We first prove
convergence of the diagonal entries $B_{ii}$ and the products $N_{ij}N_{ji}$. Note that their limits determine the normalizers in \eqref{Nlimit} and \eqref{Blimit}. Since $2\diag (B)-\Lambda$ is diagonal, \eqref{partialB} gives
\[
 \frac d{dt}\left(B_{ii}+\frac{h_{B,ii}}t+\frac{k_{B,ii}}{t^2}\right)
 =\mathcal R_{B,ii}=O(t^{-2}).
\]
Integrating this identity shows that $B_{ii}(t)$ has a limit $\beta_i$ at $t=+\infty$, with
$B_{ii}(t)-\beta_i=O(t^{-1})$.

For $i\ne j$, let us rewrite \eqref{partialN} as
\begin{equation*}
 \frac{d}{dt}\left(N_{ij}+\frac{h_{N,ij}}t+\frac{k_{N,ij}}{t^2}\right)
 =-\frac{\I(d_i-d_j)^2}{u_i-u_j}\,t
 \left(N_{ij}+\frac{h_{N,ij}}t+\frac{k_{N,ij}}{t^2}\right)+\frac{\Lambda_i-\Lambda_j-2q_{ij}}{t}N_{ij}
 +\mathcal R_{N,ij}.
\end{equation*}
The Leibniz rule gives
\begin{align*}
 &\frac{d}{dt}\left(
 \left(N_{ij}+\frac{h_{N,ij}}t+\frac{k_{N,ij}}{t^2}\right)
 \left(N_{ji}+\frac{h_{N,ji}}t+\frac{k_{N,ji}}{t^2}\right)
 \right)\\
 &\quad=\frac{\Lambda_i-\Lambda_j-2q_{ij}}{t}
 \left(
 N_{ij}\left(\frac{h_{N,ji}}t+\frac{k_{N,ji}}{t^2}\right)
 -N_{ji}\left(\frac{h_{N,ij}}t+\frac{k_{N,ij}}{t^2}\right)
 \right)+O(t^{-2})\\
 &\quad=O(t^{-2}).
\end{align*}
Since the product differs from $N_{ij}N_{ji}$ by $O(t^{-1})$,
integrating the identity shows that $N_{ij}N_{ji}$
converges as $t\to+\infty$. Consequently, we may define
\[
 q^0_{ij}=\lim_{t\to+\infty}
          \frac{N_{ij}(t)N_{ji}(t)}{\I(u_i-u_j)} \text{ for } i\ne j,
 \qquad \beta=\diag(\beta_1,\ldots,\beta_n),
\]
and obtain
\[
 q_{ij}(t)-q^0_{ij}=O(t^{-1}),\qquad
 \diag (B(t))-\beta=O(t^{-1}).
\]

Letting $t\to+\infty$ in \eqref{Lambda}, we obtain
\[
 \Lambda_i=\beta_i+\sum_{j\ne i}q^0_{ij}.
\]
\paragraph{The regularized limit of $N$.}
Set
\begin{equation}\label{muij}
 \mu_{ij}=\Lambda_i-\Lambda_j-2q^0_{ij},\qquad i\ne j.
\end{equation}
The skew-Hermitian condition gives $\mu_{ij}\in\I\mathbb R$,
and $q^0_{ji}=-q^0_{ij}$ gives $\mu_{ji}=-\mu_{ij}$.
Thus, for $t>0$,
\begin{equation}\label{scalar1}
 \left|e^{\frac{\I(d_i-d_j)^2t^2}{2(u_i-u_j)}}
       t^{-\mu_{ij}}\right|=1.
\end{equation}
Using
$\Lambda_i-\Lambda_j-2q_{ij}
=\mu_{ij}-2(q_{ij}-q^0_{ij})$
and applying the Leibniz rule to \eqref{partialN}, we get
the identity
\begin{align*}
\frac {d}{dt}\left(
 e^{\frac{\I(d_i-d_j)^2t^2}{2(u_i-u_j)}}
 t^{-\mu_{ij}}N_{ij}\right)
&=-\frac {d}{dt}\left(
 e^{\frac{\I(d_i-d_j)^2t^2}{2(u_i-u_j)}}
 t^{-\mu_{ij}}
\left(\frac{h_{N,ij}}{t}+\frac{k_{N,ij}}{t^2}\right)
 \right)\\
&\quad+
 e^{\frac{\I(d_i-d_j)^2t^2}{2(u_i-u_j)}}
 t^{-\mu_{ij}}
 \left(
 \mathcal R_{N,ij}
 -\frac{\mu_{ij}}{t}
  \left(\frac{h_{N,ij}}{t}+\frac{k_{N,ij}}{t^2}\right)
 -\frac{2(q_{ij}-q^0_{ij})}{t}N_{ij}
 \right).
\end{align*}
The three terms in the last bracket are $O(t^{-2})$, respectively
by Lemma \ref{parts}, boundedness of $h_N,k_N$, and
$q_{ij}-q^0_{ij}=O(t^{-1})$.
Since by \eqref{scalar1} the normalization factor has modulus one, integration shows that
\[
 (N_0)_{ij}:=\lim_{t\to+\infty}
 e^{\frac{\I(d_i-d_j)^2t^2}{2(u_i-u_j)}}
 t^{-\mu_{ij}}N_{ij}(t), \qquad \text{for } i\ne j,
\]
exists, with an $O(t^{-1})$ error. As for the diagonal entries, we take $(N_0)_{ii}=0$.
The normalization factors for $ij$ and $ji$ are complex conjugates
and have product one. Thus $N_0\in\mathfrak u(n)_{\off}$ and
\[
 (N_0)_{ij}(N_0)_{ji}
 =\lim_{t\to+\infty}N_{ij}(t)N_{ji}(t)
 =\I(u_i-u_j)q^0_{ij}.
\]
In particular, the exponent $\mu_{ij}$ in
\eqref{muij} is precisely the exponent in \eqref{Nlimit}.
This proves \eqref{Nlimit}.

\paragraph{The regularized limit of $B$.}
Since $2\beta-\Lambda$ is constant and skew-Hermitian,
$t^{2\beta-\Lambda}$ is unitary for $t>0$.
By \eqref{partialB}, we have
\begin{equation}\label{regB}
\begin{aligned}
 \frac {d}{dt}(t^{2\beta-\Lambda}Bt^{-(2\beta-\Lambda)})
 &=-\frac {d}{dt}\left(
 t^{2\beta-\Lambda}\left(\frac{h_B}t+\frac{k_B}{t^2}\right)t^{-(2\beta-\Lambda)}\right)\\
 &\quad+t^{2\beta-\Lambda}\left(\mathcal R_B
 +\frac{1}{t}[2\beta-2\diag (B),B]
 +\frac{1}{t}\left[2\beta-\Lambda,\frac{h_B}t+\frac{k_B}{t^2}\right]\right)t^{-(2\beta-\Lambda)}.
\end{aligned}
\end{equation}
Here recall $2(\diag (B)-\beta)=O(t^{-1})$.
Together with boundedness of $B,h_B,k_B$, this shows that all
three terms in the last bracket are $O(t^{-2})$, and conjugation
by the unitary matrix $t^{2\beta-\Lambda}$ preserves this estimate. Since
\[
 t^{2\beta-\Lambda}\left(\frac{h_B}{t}+\frac{k_B}{t^2}\right)t^{-(2\beta-\Lambda)}=O(t^{-1}),
\]
integration of \eqref{regB} gives
\[
 B_0:=\lim_{t\to+\infty}t^{2\beta-\Lambda}B(t)t^{-(2\beta-\Lambda)},\qquad
 t^{2\beta-\Lambda}B(t)t^{-(2\beta-\Lambda)}=B_0+O(t^{-1}).
\]
Moreover, since $t^{2\beta-\Lambda}$ is unitary and 
diagonal, we have 
$B_0\in\mathfrak u(n)$ and $\diag (B_0)=\beta$.
Thus $2\beta-\Lambda=2\diag (B_0)-\Lambda$, and \eqref{Blimit} follows.
It completes the proof.\qed

\section{Reference solutions and their connection formula}\label{sec:matching}
Sections \ref{secztdeg} and \ref{secijdeg} derive the limiting
systems of \eqref{zeq} near $z=0$ and near the crossing points. Section \ref{Stokesmatrices} introduces the canonical solutions and Stokes matrices of these linear systems. In the rest of the section,
we prove Theorem \ref{thm2} by comparing the canonical solutions
$F_-$ and $F_+$ of \eqref{zeq} through a chain of intermediate reference solutions. Section \ref{chain} defines the reference solutions, and Sections \ref{crossingconnections}-\ref{endpointmatching} compute the limits of the connection matrices between two consecutive reference solutions. Since the ratio $F_+^{-1}F_-$ is an ordered product of these connection
matrices, Theorem \ref{thm2} follows by taking $t\to+\infty$. 
Throughout this section, the irregular parameters $U$, $D$ are fixed and a real analytic solution $(N(t), B(t))$ is chosen. 

\subsection{The limiting system near $z=0$}\label{secztdeg}
Recall that we write the diagonal matrix
$\beta={\rm diag}(B_0)$, with
$\beta_k=(B_0)_{kk}$, and
by the proof of Theorem \ref{thm1},
${\rm diag}(B(t))=\beta+O(t^{-1})$.

Introduce the change of variables $\zeta=zt,t=t$ and the transform
$\widetilde F(\zeta,t)=t^{2\beta-\Lambda}F(\zeta/t,t)$. Then the system \eqref{zeq}-\eqref{teq} becomes
\begin{align}\label{zetaeq}
 \frac{\partial \widetilde F}{\partial\zeta}
 &=\left(D+\frac{U\zeta}{t^2}
 +\frac{t^{2\beta-\Lambda}Nt^{-(2\beta-\Lambda)}}{t}+\frac{t^{2\beta-\Lambda}Bt^{-(2\beta-\Lambda)}}{\zeta}\right)\widetilde F,\\
 \frac{\partial\widetilde F}{\partial t}
 &=\left(t^{2\beta-\Lambda}\Big(\T N-\frac Bt
       -\frac{N\zeta}{t^2}\Big)t^{-(2\beta-\Lambda)}-\frac{U\zeta^2}{t^3}
       +\frac{2\beta-\Lambda}{t}\right)\widetilde F.
\end{align}
By \eqref{Blimit} and boundedness of $N(t)$, as
$t\to+\infty$ along the positive real axis, the coefficient matrix
of the first equation converges to $D+B_0/\zeta$,
uniformly for $\zeta$ in compact subsets of
$\mathbb C\setminus\{0\}$. The limiting equation becomes an $n\times n$ linear system  
\begin{equation}\label{Yeq}
 \frac{d Y}{d\zeta }=\left(D+\frac{B_0}{\zeta}\right)Y.
\end{equation}

\subsection{The limiting systems near the crossing points}\label{secijdeg}

Fix indices $1\le i<j\le n$ and introduce the crossing point $z_{ij}=-t\frac{d_i-d_j}{u_i-u_j}$. In terms of the variables
\[
 \eta=z-z_{ij}=z+t\frac{d_i-d_j}{u_i-u_j},\qquad t=t
\]
the system \eqref{zeq}-\eqref{teq} becomes
\begin{align}\label{Fetaeq}
\frac{\partial F}{\partial\eta}
 &=\left(tV+U\eta+N
   +\frac{B}{\eta-t\frac{d_i-d_j}{u_i-u_j}}\right)\cdot F,
 \\ \label{Fteq}
\frac{\partial F}{\partial t}
 &=\left(
 V\eta
 -t\frac{d_i-d_j}{u_i-u_j}
      \Big(2D-\frac{d_i-d_j}{u_i-u_j}U\Big)
 -\frac{d_i-d_j}{u_i-u_j}N+\T N
 -\frac{\frac{d_i-d_j}{u_i-u_j}B}
       {\eta-t\frac{d_i-d_j}{u_i-u_j}}
 \right)\cdot F.
\end{align}
where $V={\rm diag}(V_1,\ldots,V_n)$ is the diagonal matrix 
\[V=D-\frac{d_i-d_j}{u_i-u_j}U.\]

By definition, $V_i=V_j$. By the generic condition \eqref{separation}, no other pair of diagonal entries of $V$ coincides. We introduce the following transformation that
normalizes the $(ij)$ and $(ji)$ entries of $N$ and reduces the other off-diagonal terms to a small remainder
\begin{equation}\label{intrans}
 F(\eta,t)
 =R^{(ij)}(t)
  \left(1+{H^{(ij)}(t)}{t^{-1}}\right)
  e^{tV\eta}\cdot \widetilde F(\eta,t).
\end{equation}
Here the superscript $(ij)$ indicates the chosen pair $i<j$, and
\begin{itemize}
\item $R^{(ij)}(t)$ is the $n\times n$ diagonal matrix  
\begin{equation}\label{Rdefi}
 R^{(ij)}(t)
={\rm e}^{\frac{Uz_{ij}^{2}}{2}+tDz_{ij}}
 t^{\Lambda-q^0_{ij}(E_{ii}-E_{jj})}=\exp\Big(-\frac{t^2}{2}\frac{d_i-d_j}{u_i-u_j}(D+V)\Big)
 t^{\Lambda-q^0_{ij}(E_{ii}-E_{jj})}.
\end{equation}

 \item $H^{(ij)}(t)$ is the $n\times n$ matrix with entries
\begin{equation}\label{Hij}
H^{(ij)}_{ab}
=\begin{cases}
-\dfrac{\bigl((R^{(ij)})^{-1}N R^{(ij)}\bigr)_{ab}}{V_a-V_b},
   & V_a\ne V_b,\\[4pt]
0, & V_a=V_b.
\end{cases}
\end{equation}
\end{itemize}

\begin{lem}\label{inner}
Under the transformation \eqref{intrans}, equation
\eqref{Fetaeq} becomes
\begin{equation}\label{inequation}
 \frac{\partial\widetilde F}{\partial\eta}
 =\left(U\eta+(N_0)_{ij}E_{ij}+(N_0)_{ji}E_{ji}
 +\mathcal M^{(ij)}(\eta,t)\right)\widetilde F.
\end{equation}
For sufficiently large $t$ and real $\eta$ with $|\eta|\le |z_{ij}|/2$,
\[
 \norm{\mathcal M^{(ij)}(\eta,t)}\le \frac{C(1+|\eta|)}{t},
\]
where $C$ is independent of $t$ and $\eta$.
Consequently, the coefficient of \eqref{inequation} converges
uniformly on compact real $\eta$ intervals as $t\to+\infty$
to that of the $n\times n$ system
\begin{equation}\label{nijeq}
 \frac{dK^{(ij)}}{d\eta}
 =\left(U\eta+(N_0)_{ij}E_{ij}+(N_0)_{ji}E_{ji}\right)K^{(ij)}.
\end{equation}
\end{lem}

\begin{proof}
Set $\widetilde N=(R^{(ij)})^{-1}NR^{(ij)}$ and for a matrix $X$ write
$X_{(ij)}=X_{ij}E_{ij}+X_{ji}E_{ji}$.
By \eqref{Hij}, $[V,H^{(ij)}]=-\widetilde N+\widetilde N_{(ij)}$.
Also note that $[V,U]=[V,\widetilde N_{(ij)}]=0$.
Substituting \eqref{intrans} into \eqref{Fetaeq} gives
\eqref{inequation}, with
\begin{align*}
 \mathcal M^{(ij)}(\eta,t)
 &=\widetilde N_{(ij)}-(N_0)_{(ij)}\\
 +&e^{-tV\eta}\Big(1+\frac{H^{(ij)}}{t}\Big)^{-1}
 \left(
 \frac{\eta[U,H^{(ij)}]+\widetilde N H^{(ij)}
       -H^{(ij)}\widetilde N_{(ij)}}{t}
 +\frac{(R^{(ij)})^{-1}BR^{(ij)}(1+H^{(ij)}/t)}
       {z_{ij}+\eta}
 \right)e^{tV\eta}.
\end{align*}
By \eqref{Rdefi} and \eqref{Nlimit},
\[
 \widetilde N_{ij}
 =\frac{(R^{(ij)})_{jj}}{(R^{(ij)})_{ii}}N_{ij}
 =\exp\Big(\frac{\I(d_i-d_j)^2t^2}{2(u_i-u_j)}\Big)
 t^{-\mu_{ij}}N_{ij}
 =(N_0)_{ij}+O(t^{-1}).
\]
The same estimate holds for the $(ji)$ entry, so
$\widetilde N_{(ij)}-(N_0)_{(ij)}=O(t^{-1})$.

Since $R^{(ij)}$ is unitary, $\widetilde N$ is bounded.
The nonzero denominators in \eqref{Hij} are fixed, so $H^{(ij)}$
is also bounded. The first fraction in the last bracket of the expression of $\mathcal M^{(ij)}(\eta,t)$ is therefore
$O((1+|\eta|)/t)$.

For $|\eta|\le |z_{ij}|/2$, we have $|z_{ij}+\eta|\ge |z_{ij}|/2$.
Since $|z_{ij}|$ is a fixed positive multiple of $t$, the second
fraction in the last bracket of $\mathcal M^{(ij)}(\eta,t)$ is $O(t^{-1})$. The matrices $e^{\pm tV\eta}$ are unitary
for real $\eta$, and $(1+H^{(ij)}/t)^{-1}$ is bounded for
sufficiently large $t$. These estimates give the required bound.
\end{proof}

\subsection{Canonical solutions and Stokes matrices}\label{Stokesmatrices}
The content in this subsection is standard in the literature of meromorphic linear ordinary differential equations, see e.g., \cite{Balser, LR, Wa}.

For any fixed $t$, since $U=\I\diag(u_1,\ldots,u_n)$ has distinct eigenvalues, the linear equation
\eqref{zeq} has a unique formal fundamental solution of the form
\begin{equation}
 \widehat F(z,t)
 =\widehat H(z,t)\cdot e^{\frac{Uz^2}{2}+tDz}z^{\Lambda}, \qquad \widehat H(z,t)=1+H_1(t)z^{-1}+H_2(t)z^{-2}+\cdots
\end{equation}
where $\Lambda$ is the diagonal matrix defined as in \eqref{Lambda}. In general, the radius of convergence of the matrix formal power series $\widehat H(z,t)$ in $z^{-1}$ is zero. However, standard
summation procedures produce holomorphic matrix functions
on certain sectors around $z=\infty$, having $\widehat H(z,t)$ as their
asymptotic expansion. Multiplied by the singular term
$e^{Uz^2/2+tDz}z^\Lambda$, these holomorphic matrix functions then give fundamental solutions
of \eqref{zeq}. These sectorial solutions agree when the summation direction varies
between two consecutive anti-Stokes directions.
The fundamental solutions obtained on the two sides of an
anti-Stokes direction are related by a Stokes matrix. 
We now describe these canonical solutions and Stokes matrices for \eqref{zeq}.

Since the leading term $U$ has purely imaginary eigenvalues, the anti-Stokes directions of the $z$-equation \eqref{zeq} are
$\arg z=\pi/4+j\pi/2$, $j\in\mathbb Z$. For any $j$ define the sector
\[
 \Sect_j=\left\{z:
 \left|\arg z-\frac{j\pi}{2}\right|<\frac{\pi}{2}\right\}
\]
on the universal cover of $\mathbb{C}^*$.

\begin{pro}
\label{CanSol}
For any $j\in\mathbb Z$, there exists a  unique (therefore canonical) fundamental solution $F_{\infty,j}$ of the $z$-equation \eqref{zeq} with the prescribed asymptotics
\begin{equation}\label{asymcond}
 F_{\infty,j}(z,t)\cdot e^{-\frac{Uz^2}{2}-tDz}z^{-\Lambda} \sim \widehat H(z,t)=1+O(z^{-1}), \qquad \text{
as $z\to\infty$ within $\Sect_j$.}
\end{equation}
\end{pro}
\begin{defi}
For any $t$ and $j\in\mathbb{Z}$, the Stokes matrix $S_j(U,tD;N(t),B(t))\in {\rm GL}_n$ of the $z$-equation \eqref{zeq} is defined by
\begin{equation}\label{defS}
 F_{\infty,j+1}(z,t)=F_{\infty,j}(z,t)\cdot S_j.
\end{equation}
\end{defi}

\begin{lem}
The Stokes matrices satisfy the following properties:
\begin{itemize}
    \item[(a).] For $k\in\mathbb{Z}$,
$S_{2k+1}$ is upper unitriangular and $S_{2k}$ is
lower unitriangular;
\item[(b).] For $j\in\mathbb{Z}$, we have $S_{j+4}=e^{-2\pi\I\Lambda}S_j e^{2\pi\I\Lambda}$;
\item[(c).] For real $t$ we have 
\begin{equation}\label{Sconj}
S_2=e^{-2\pi\I\Lambda} S_1^\dagger e^{2\pi\I\Lambda},\qquad
 S_3=e^{-2\pi\I\Lambda} S_0^\dagger e^{2\pi\I\Lambda}.
\end{equation}
\end{itemize}
\end{lem}
\begin{proof}
Part $(a)$ follows from the asymptotic condition \eqref{asymcond}. For part $(b)$, just note that the solution $F_{\infty,j+4}$ is normalized
on the next sheet, i.e.,
\begin{equation}\label{z2pii}
 F_{\infty,j+4}(ze^{2\pi\I},t)=F_{\infty,j}(z,t)\cdot e^{2\pi\I\Lambda}.
\end{equation}
For part $(c)$, for real $t$, the skew-Hermitian coefficient matrices give the complex conjugation transpose
\[
 \left(U\bar z+tD+N(t)+\frac{B(t)}{\bar z}\right)^\dagger
 =-\left(Uz+tD+N(t)+\frac{B(t)}{z}\right).
\]
Hence $(F(\bar z,t)^\dagger)^{-1}$ solves the same
equation as $F(z,t)$. The
uniqueness of the canonical solutions gives
\begin{equation*}
 \bigl(F_{\infty,j}(\bar z,t)^\dagger\bigr)^{-1}
 =F_{\infty,-j}(z,t),\qquad j\in\mathbb Z.
\end{equation*}
Together with \eqref{defS} and \eqref{z2pii}, it proves \eqref{Sconj}.
\end{proof}
 
Thus $\Lambda,S_0,S_1$ determine all Stokes matrices. Set $F_+=F_{\infty,0}$ and $F_-=F_{\infty,2}$. Then
\begin{equation*}
 F_-(z,t)=F_+(z,t)\cdot S_0S_1.
\end{equation*}

It follows from the theory of isomonodromy deformation \cite{JMU} that 
\begin{pro}
For any $j\in\mathbb{Z}$, the solution $F_{\infty,j}(z,t)$ satisfies the time $t$-equation \eqref{teq}. Hence the Stokes matrices $S_{j}(U,tD;N(t),B(t))$ are independent of $t$. 
\end{pro}

Similarly, we introduce Stokes matrices of the two limiting linear systems. The limiting system \eqref{Yeq} has a simple pole at $\zeta=0$ and a pole of order
two at $\zeta=\infty$, with anti-Stokes directions
$\arg\zeta=\pi/2+j\pi$.
\begin{defi}
Let $Y_+$ and $Y_-$ be its canonical fundamental solutions with the prescribed asymptotics
\[
 Y_\pm(\zeta)\cdot  e^{-D\zeta}\zeta^{-\beta}=1+O(\zeta^{-1}),
\]
as $\zeta\to \infty$ within the sectors $-\pi<{\rm arg}(\zeta)<\pi$ and $0<{\rm arg}(\zeta)<2\pi$ respectively.
Define
the lower unitriangular Stokes matrix by
\begin{equation}
 Y_-=Y_+S_{\out}(D,B_0).
\end{equation}
\end{defi}

For each pair $1\le i<j\le n$, we take the $n\times n$ linear system \eqref{nijeq}, i.e., 
\begin{equation}
 \frac{dK^{(ij)}}{d\eta}
 =\left(U\eta+(N_0)_{ij}E_{ij}+(N_0)_{ji}E_{ji}\right)\cdot K^{(ij)}.
\end{equation}
It has a pole of order three at $\eta=\infty$ and has anti-Stokes directions
$\arg\eta=\pi/4+k\pi/2$.
\begin{defi}
Let $K_k^{(ij)}$ be its canonical fundamental solution normalized by
\begin{equation}\label{Kasym}
 K_k^{(ij)}(\eta)e^{-U\eta^2/2}
 \eta^{-q^0_{ij}(E_{ii}-E_{jj})}
 =1+O(\eta^{-1}),\qquad
 \eta\to\infty\ \text{within }\Sect_k.
\end{equation}
Define its four Stokes matrices $S_k^{(ij)}\in{\rm GL}_n$ by
\begin{equation}
K_{k+1}^{(ij)}=K_k^{(ij)}S_k^{(ij)},\qquad k=0,1,2,3.
\end{equation}
\end{defi}
Note that the system \eqref{nijeq} decouples into $n-2$ scalar equations and a $2\times 2$ block. Consequently, for $a\ne i,j$, the $(aa)$ entry of $K_k^{(ij)}$ is
$e^{\I u_a\eta^2/2}$, and all off-diagonal entries, except the $(ij),(ji)$ positions, are zero.
Accordingly, the matrices $S_0^{(ij)},S_2^{(ij)}$ differ from the identity matrix only
in position $(j,i)$, and $S_1^{(ij)},S_3^{(ij)}$ differ from
the identity only in position $(i,j)$, see \eqref{Sentry} for their expressions.
Set the solutions $K_+^{(ij)}=K_0^{(ij)}$ and $K_-^{(ij)}=K_2^{(ij)}$, 
then their connection
matrix is
\begin{equation*}
 K_-^{(ij)}=K_+^{(ij)}\cdot S_0^{(ij)}S_1^{(ij)}.
\end{equation*}

\subsection{A chain of intermediate reference solutions}\label{chain}

As in the introduction, we order the $m=n(n-1)/2$ pairs
$(p_j,q_j)$, with $p_j<q_j$, such that \eqref{orderpq} holds.
Set
\[
 L(t)=t^{1/3},\qquad
 z_j=z_{p_jq_j}=-t\frac{d_{p_j}-d_{q_j}}{u_{p_j}-u_{q_j}},\qquad
 z_j^\pm=z_j\pm L.
\]
For sufficiently large $t$, the intervals $[z_j^-,z_j^+]$ are
disjoint and ordered from left to right on the negative real axis. 

\begin{defi}\label{references}
For each fixed sufficiently large $t$, let $\mathcal F_j^\pm(z,t)$
be the fundamental solutions of \eqref{zeq} determined by the initial condition at $z=z_j^\pm$
\begin{equation}\label{FbyPhi}
 \mathcal F_j^\pm(z_j^\pm,t)=\Phi(z_j^\pm,t),
\end{equation}
where $\Phi$ is the diagonal matrix function with entries
\begin{equation}\label{Phi}
 \Phi_{kk}(z,t)=e^{\I u_kz^2/2+\I t d_kz}
 z^{\beta_k}\prod_{l\ne k}
 \Big(z+t\frac{d_k-d_l}{u_k-u_l}\Big)^{q^0_{kl}}.
\end{equation}
Here recall that $\beta_k=(B_0)_{kk}$ and $q^0_{kl}=\frac{(N_0)_{kl}(N_0)_{lk}}{\I(u_k-u_l)},$ and we use
$\log x=\log|x|+\pi\I$ if $x<0$.
\end{defi}
Since $\beta_k+\sum_{l\ne k}q^0_{kl}=\Lambda_k$, for fixed $t$,
\begin{equation}\label{Phiinfinity}
\Phi(z,t)=\bigl(1+O(z^{-1})\bigr)
 e^{\frac{Uz^2}{2}+tDz}z^{\Lambda},\qquad \text{as } z\to\pm\infty \text{ along the real axis}.
\end{equation} 
Moreover, for real $z$, since $\beta_k,q^0_{kl}$ are purely
imaginary,
\begin{equation}\label{Phibound}
\max\left\{|\Phi_{kk}(z,t)|,\,
             |\Phi_{kk}(z,t)|^{-1}\right\}
 \le \exp\Big(\pi\Big(
 |\beta_k|+\sum_{l\ne k}|q^0_{kl}|\Big)\Big), \qquad {k=1,\ldots,n}.
\end{equation}
Thus $\Phi(z,t)$ and $\Phi(z,t)^{-1}$ are
uniformly bounded for $t>0$ and real $z$ for which $\Phi(z,t)$ is defined.
\begin{defi}
For sufficiently large $t$, the connection matrices
$M_0(t)$, $W_a(t)$, $M_a(t)$ and $M_+(t)$ are defined by
\begin{equation}\label{refconnection}
\begin{aligned}
 M_0(t)&=\mathcal F_1^-(z,t)^{-1}F_-(z,t), \qquad M_+(t)=F_+(z,t)^{-1}\mathcal F_m^+(z,t),\\
 W_a(t)&=\mathcal F_a^+(z,t)^{-1}\mathcal F_a^-(z,t), \ \ 1\le a\le m \\
 M_a(t)&=\mathcal F_{a+1}^-(z,t)^{-1}\mathcal F_a^+(z,t), \ \ 1\le a<m.
\end{aligned}
\end{equation}
\end{defi}

These matrices are independent of $z$, since the fundamental
solutions satisfy the same equation \eqref{zeq}. We write the Stokes matrices as an ordered product of the connection matrices 
\begin{equation}\label{orderpro}
S_0S_1=(F_+)^{-1}F_-
=
M_+(t)W_m(t)M_{m-1}(t)W_{m-1}(t)
\cdots M_1(t)W_1(t)M_0(t).
\end{equation}

\begin{figure}[htbp]
 \centering
 \resizebox{\linewidth}{!}{%
 \begin{tikzpicture}[
  x=1cm,y=1cm,>=stealth,
  reference/.style={inner sep=2pt},
  connection/.style={line width=.8pt,<-},
  labelabove/.style={above=3pt,inner sep=2pt,fill=white},
  ticklabel/.style={anchor=north}
 ]
 \node[reference] (minus) at (.55,0) {$F_-$};
 \node[reference] (firstminus) at (2.8,0) {$\mathcal F_1^-$};
 \node[reference] (firstplus) at (5.1,0) {$\mathcal F_1^+$};
 \node[reference] (middle) at (8.0,0) {$\cdots$};
 \node[reference] (lastminus) at (10.7,0) {$\mathcal F_m^-$};
 \node[reference] (lastplus) at (13,0) {$\mathcal F_m^+$};
 \node[reference] (plus) at (19.6,0) {$F_+$};
 \draw[connection,blue!65!black] (minus.east)--
  node[labelabove] {$M_0(t)$} (firstminus.west);
 \draw[connection,teal!80!black] (firstminus.east)--
  node[labelabove] {$W_1(t)$} (firstplus.west);
 \draw[connection,blue!65!black] (firstplus.east)--
  node[labelabove] {$M_1(t)$} (middle.west);
 \draw[connection,blue!65!black] (middle.east)--(lastminus.west);
 \draw[connection,teal!80!black] (lastminus.east)--
  node[labelabove] {$W_m(t)$} (lastplus.west);
 \draw[connection,black!75] (lastplus.north)
  to[out=35,in=145]
  node[above=3pt,fill=white,inner sep=2pt]
  {$M_+(t)$} (plus.north);
 \node[black] at (16.4,-.45)
  {$Y_-=Y_+S_{\out}$};
 \fill[teal!9] (2.8,-1.52) rectangle (5.1,-1.18);
 \fill[teal!9] (10.7,-1.52) rectangle (13,-1.18);
 \fill[black!5] (16.3,-1.52) rectangle (18.3,-1.18);
 \draw[->,line width=.8pt] (.05,-1.35)--(20.5,-1.35)
  node[right] {$z$};
 \foreach \x in {2.8,5.1,10.7,13}{
  \draw[gray!55,densely dotted] (\x,-.35)--(\x,-1.28);
  \fill (\x,-1.35) circle (1.6pt);
 }
 \foreach \x in {3.95,11.85}{
  \draw[teal!80!black,fill=white,line width=.8pt]
   (\x,-1.35) circle (2pt);
  \draw[teal!60,densely dotted] (\x,-1.47)--(\x,-2.35);
 }
 \draw[black!75,line width=.8pt] (16.3,-1.2)--(16.3,-1.5);
 \draw[black!75,line width=.8pt] (18.3,-1.2)--(18.3,-1.5);
 \draw[fill=white,line width=.8pt] (17.3,-1.35) circle (2.2pt);
 \node[ticklabel] at (.55,-1.68) {$-\infty$};
 \node[ticklabel] at (2.8,-1.68) {$z_1^-$};
 \node[ticklabel] at (5.1,-1.68) {$z_1^+$};
 \node[ticklabel] at (10.7,-1.68) {$z_m^-$};
 \node[ticklabel] at (13,-1.68) {$z_m^+$};
 \node[ticklabel] at (16.3,-1.68) {$-L/t$};
 \node[ticklabel] at (17.3,-1.68) {$0$};
 \node[ticklabel] at (18.3,-1.68) {$L/t$};
 \node[ticklabel] at (19.6,-1.68) {$+\infty$};
 \node[teal!80!black] at (3.95,-2.7)
  {$-t\dfrac{d_{p_1}-d_{q_1}}{u_{p_1}-u_{q_1}}$};
 \node[teal!80!black] at (11.85,-2.7)
  {$-t\dfrac{d_{p_m}-d_{q_m}}{u_{p_m}-u_{q_m}}$};
 \node[blue!65!black] at (8,-1.95) {comparison by $\Phi$};
 \end{tikzpicture}%
 }
 \caption{Comparison of $F_-$ and $F_+$ through the reference solutions. The $\mathcal F_a^\pm$ are solutions of \eqref{zeq} with initial values $\Phi(z_a^\pm,t)$ at the reference points $z_a^{\pm}$.
 The arrows labelled $W_a(t)$ and $M_a(t)$ mean
 $\mathcal F_a^-=\mathcal F_a^+W_a(t)$ and
 $\mathcal F_a^+=\mathcal F_{a+1}^-M_a(t)$, respectively.
 The endpoint arrows indicate comparison with the two canonical
 solutions $F_\pm$.}
\end{figure}
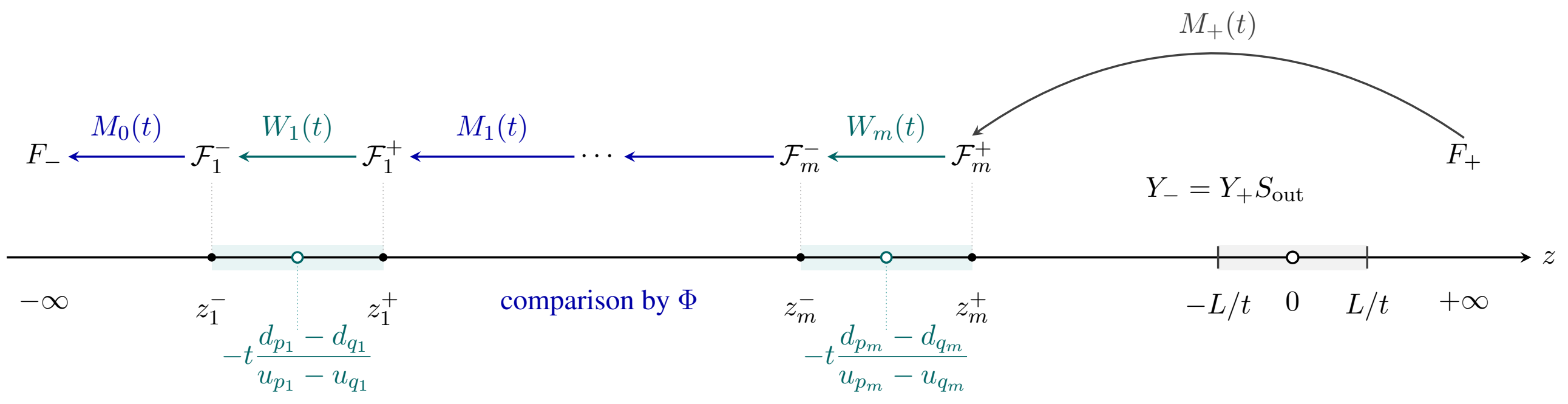

In the next three
subsections, we give the limits of the connection matrices as $t\to+\infty$,
\begin{align*}
M_0(t)
&\longrightarrow 1,\qquad
W_a(t)
\longrightarrow
T^{(p_aq_a)}=(D^{(p_aq_a)})^{-1}\cdot
 S_0^{(p_aq_a)}S_1^{(p_aq_a)}\cdot D^{(p_aq_a)},\\
M_a(t)
&\longrightarrow
1,\qquad
M_+(t)
\longrightarrow
D_{\out}^{-1}S_{\out}(D,B_0)D_{\out}.
\end{align*}
Taking limits in the ordered product \eqref{orderpro} then proves
\eqref{factS}.

\subsection{Connection matrices across the crossing intervals}
\label{crossingconnections}
Recall that the matrices $D^{(ij)}$ and $T^{(ij)}$
are defined in  \eqref{Dmatrix} and \eqref{Tmatrix}.

\begin{pro}\label{proWlimit}
Fix $(i,j)=(p_a,q_a)$.
Apply the transformation \eqref{intrans} to the reference
solutions $\mathcal F_a^\pm$ in Definition \ref{references}, and
denote the transformed solutions by $\widetilde{\mathcal F}_a^\pm$,
\begin{equation}\label{normalizedref}
 \widetilde{\mathcal F}_a^\pm(\eta,t)
 =e^{-tV\eta}\left(1+{H^{(ij)}}/t\right)^{-1}
 (R^{(ij)})^{-1}\mathcal F_a^\pm(z_a+\eta,t).
\end{equation}
For $L(t)=t^{1/3}$ we have
\begin{equation}\label{innerconve}
 \sup_{\eta\in\mathbb{R}, |\eta|\le L}
 \norm{\widetilde{\mathcal F}_a^\pm(\eta,t)
       -K_\pm^{(ij)}(\eta)D^{(ij)}}=O(t^{-1/3}).
\end{equation}
Consequently,
\begin{equation}\label{Wlimit}
 W_a(t)=(\mathcal F_a^+)^{-1}\mathcal F_a^-
 =T^{(ij)}+O(t^{-1/3}).
\end{equation}
\end{pro}

\begin{proof}
We first compare $\widetilde{\mathcal F}_a^\pm(\eta,t)$ and $K^{(ij)}_\pm(\eta)D^{(ij)}$ at the endpoints $\eta=\pm L$. By \eqref{FbyPhi} and \eqref{normalizedref},
\[
 \widetilde{\mathcal F}_a^\pm(\pm L,t)
 =e^{\mp tVL}\left(1+{H^{(ij)}}/t\right)^{-1}
  (R^{(ij)})^{-1}\Phi(z_a\pm L,t).
\]
Substituting $z=z_a\pm L$ into \eqref{Phi}, and using
the definitions of $R^{(ij)}$ and $D^{(ij)}$, we get
\begin{equation*}
 \widetilde{\mathcal F}_a^\pm(\pm L,t)
 =e^{\mp tVL}\left(1+{H^{(ij)}}/t\right)^{-1}
       e^{\pm tVL}e^{UL^2/2}
   (\pm L)^{q^0_{ij}(E_{ii}-E_{jj})}
   D^{(ij)}\bigl(1+O(L/t)\bigr).
\end{equation*}
The product of the first three factors is $1 + O(t^{-1})$, because $H^{(ij)}$ is bounded and $e^{\pm tVL}$ is unitary. Since the other factors are bounded, we obtain
\[
 \widetilde{\mathcal F}_a^\pm(\pm L,t)
 =e^{UL^2/2}(\pm L)^{q^0_{ij}(E_{ii}-E_{jj})}
  D^{(ij)}+O(L/t).
\]

On the other hand, \eqref{Kasym} gives
\[
 K_\pm^{(ij)}(\pm L)e^{-UL^2/2}
 (\pm L)^{-q^0_{ij}(E_{ii}-E_{jj})}
 =1+O(L^{-1}).
\]
Comparing these two expressions, we obtain
\begin{equation}\label{compFK}
 \widetilde{\mathcal F}_a^\pm(\pm L,t)=K^{(ij)}_\pm(\pm L)D^{(ij)}
   +O(L^{-1}+L/t)=K^{(ij)}_\pm(\pm L)D^{(ij)}+O(L^{-1}),
\end{equation}
where the last equality uses $L=t^{1/3}$.

Since the coefficient of \eqref{zeq} is skew-Hermitian
on $[z_a^-,z_a^+]$, its propagator is unitary, hence
\eqref{FbyPhi} gives
\[
 \|\mathcal F_a^\pm(z,t)\|
 =\|\Phi(z_a^\pm,t)\|\le C,
 \qquad z\in[z_a^-,z_a^+].
\]
The transformation \eqref{normalizedref} therefore yields
\begin{equation}\label{Fabound}
 \sup_{|\eta|\le L}
 \|\widetilde{\mathcal F}_a^\pm(\eta,t)\|\le C.
\end{equation}

By Lemma \ref{inner}, the transformed reference solution satisfies \eqref{inequation}. The variation of constants formula gives
\begin{align*}
 &\widetilde{\mathcal F}_a^\pm(\eta,t)
       -K^{(ij)}_\pm(\eta)D^{(ij)}\\
       =&K^{(ij)}_\pm(\eta)K^{(ij)}_\pm(\pm L)^{-1}
 \left(\widetilde{\mathcal F}_a^\pm(\pm L,t)
       -K^{(ij)}_\pm(\pm L)D^{(ij)}\right)+\int_{\pm L}^{\eta}
 K^{(ij)}_\pm(\eta)K^{(ij)}_\pm(s)^{-1}
 \mathcal M^{(ij)}(s,t)
 \widetilde{\mathcal F}_a^\pm(s,t)\,ds.
\end{align*}
The coefficient of the limiting system is skew-Hermitian,
so $K^{(ij)}_\pm(\eta)K^{(ij)}_\pm(s)^{-1}$ is unitary for real $\eta,s$. Since $L=t^{1/3}\le |z_{ij}|/2$ for sufficiently large
$t$, the  estimate in Lemma \ref{inner} applies on
$[-L,L]$. Using the endpoint estimate \eqref{compFK} and the
bound \eqref{Fabound}, we obtain
\begin{align}
 \sup_{|\eta|\le L}
 \left\|\widetilde{\mathcal F}_a^\pm(\eta,t)
       -K^{(ij)}_\pm(\eta)D^{(ij)}\right\|
 &\le \frac{C}{L}
      +C\int_{-L}^{L}\|\mathcal M^{(ij)}(s,t)\|\,ds
      \nonumber\\
 &\le \frac{C}{L}+\frac{C}{t}\int_{-L}^{L}(1+|s|)\,ds
 =\frac{C}{L}+\frac{C(2L+L^2)}{t}.
 \label{FKD}
\end{align}
With $L=t^{1/3}$, both terms on the right are $O(t^{-1/3})$.

Note that the common left factor in \eqref{normalizedref} cancels in the ratio. For real $\eta$ with $|\eta|\le L$ and $z=z_a+\eta$, we have
\[
\begin{aligned}
 W_a(t)
 &=\mathcal F_a^+(z,t)^{-1}\mathcal F_a^-(z,t)\\
 &=\widetilde{\mathcal F}_a^+(\eta,t)^{-1}
   \widetilde{\mathcal F}_a^-(\eta,t)\\
 &=(D^{(ij)})^{-1}K^{(ij)}_+(\eta)^{-1}K^{(ij)}_-(\eta)D^{(ij)}
   +O(t^{-1/3})\\
 &=(D^{(ij)})^{-1}S_0^{(ij)}S_1^{(ij)}D^{(ij)}
   +O(t^{-1/3}).
\end{aligned}
\]
Here the third identity uses the fact that $K_\pm^{(ij)}(\eta)$ and their inverses are bounded for real $\eta$. It proves \eqref{Wlimit}.
\end{proof}

\subsection{Connection matrices between consecutive crossing intervals}

We show that the two reference solutions $\mathcal F_j^+$ and $\mathcal F_{j+1}^-$ normalized at the ends of
$[z_j^+,z_{j+1}^-]$ are both approximated by $\Phi$ throughout
that interval. This gives the following estimate for their connection matrix $M_j(t)$.

\begin{pro}\label{compPhiF}
Let $L=t^{1/3}$. For $1\le j<m$,
\begin{equation}\label{PhiFcomp}
 \sup_{z\in[z_j^+,z_{j+1}^-]}
 \left(\|\Phi^{-1}\mathcal F_j^+-1\|
       +\|\Phi^{-1}\mathcal F_{j+1}^--1\|\right)
 =O(L^{-1}+t^{-1}\log t).
\end{equation}
Moreover,
\begin{equation}\label{PhiEndpointComp}
\begin{aligned}
 \Phi(z,t)^{-1}F_-(z,t)
 &=1+O(L^{-1}+t^{-1}\log t),
 &&z\in(-\infty,z_1^-],\\
 \Phi(z,t)^{-1}\mathcal F_m^+(z,t)
 &=1+O(L^{-1}+t^{-1}\log t),
 &&z\in[z_m^+,-L/t],\\
 \Phi(z,t)^{-1}F_+(z,t)
 &=1+O(L^{-1}+t^{-1}\log t),
 &&z\in[L/t,+\infty).
\end{aligned}
\end{equation}
These estimates are uniform on the indicated intervals.
Consequently,
\begin{equation}\label{Mlimit}
 M_j(t)=1+O(L^{-1}+t^{-1}\log t),\qquad 1\le j<m,
\end{equation}
and
\begin{equation}\label{M0limit}
 M_0(t)=1+O(t^{-1/3}).
\end{equation}
\end{pro}

\begin{proof}
Let $I_t$ be any one of the intervals in
\eqref{PhiFcomp}-\eqref{PhiEndpointComp}.
We compare the solutions in the statement with $Q\Phi$, where $Q$
is a unitary matrix diagonalizing the coefficient matrix of \eqref{zeq} for $z\in I_t$.
The main estimate that we will use is
\begin{equation}\label{QPhiest}
 \int_{I_t}
 \left\|\partial_z(Q\Phi)-(Uz+tD+N+B/z)Q\Phi\right\|\,dz
 =O(L^{-1}+t^{-1}\log t).
\end{equation}

Let us first introduce the matrix function $Q$.
We regard the coefficient of \eqref{zeq} as a perturbation
of the diagonal matrix $Uz+tD$ by $N+B/z$. Let $g(z)$ be the minimum eigenvalue gap of $Uz+tD$
\[
 g(z)=\min_{k<l}|(u_k-u_l)z+t(d_k-d_l)|.
\]
Using the boundedness of $N,B$ and the definition of $I_t$,
we have
\[
 \sup_{z\in I_t}\frac{\|N+B/z\|}{g(z)}
 \le C\sup_{z\in I_t}\frac{1+|z|^{-1}}{g(z)}
 =O(L^{-1}).
\]
For each sufficiently large $t$, standard perturbation
estimates for skew-Hermitian matrices give a smooth unitary
matrix $Q(\,\cdot\,,t)$ on the real interval $I_t$, with diagonal entries
$Q_{kk}>0$, such that
\begin{equation}\label{Qdefi}
 Q^{-1}\left(Uz+tD+N+\frac Bz\right)Q
 ={\rm diag}(\lambda_1,\ldots,\lambda_n),
 \qquad
 \|Q-1\|\le C\frac{\|N+B/z\|}{g}\le C\frac{1+|z|^{-1}}{g}.
\end{equation}
In particular, $Q=1+O(L^{-1})$ uniformly on $I_t$.
By \eqref{Qdefi},
\[
 \partial_z(Q\Phi)-(Uz+tD+N+B/z)Q\Phi
 =\partial_zQ\cdot\Phi+Q\left(\partial_z\Phi-
 \operatorname{diag}(\lambda_1,\ldots,\lambda_n)\Phi\right).
\]
Since $Q$ is unitary for $z\in I_t$ and $\Phi$ is bounded by \eqref{Phibound}, to prove \eqref{QPhiest}
it suffices to show that
\begin{equation}\label{Qphilambda}
 \int_{I_t}\left(\|\partial_zQ\|+
 \sum_k\left|\lambda_k-
 \frac{\partial_z\Phi_{kk}}{\Phi_{kk}}\right|\right)\,dz
 =O(L^{-1}+t^{-1}\log t).
\end{equation}

To estimate $\partial_z Q$, we differentiate \eqref{Qdefi} with respect to $z$ at fixed $t$
and take the $(lk)$ entry for $l\ne k$, 
\[
 (Q^{-1}\partial_z Q)_{lk}
 =\frac{\bigl(Q^{-1}(U-B/z^2)Q\bigr)_{lk}}
        {\lambda_k-\lambda_l}.
\]
Since $U$ is diagonal, \eqref{Qdefi} gives
$(Q^{-1}UQ)_{lk}=O((1+|z|^{-1})/g)$ for $l\ne k$. Moreover,
$\left\|B/z^2\right\|\le C|z|^{-2}$ and $|\lambda_k-\lambda_l|\ge g/2$
for sufficiently large $t$. Thus
\[
 (Q^{-1}\partial_z Q)_{lk}=O\left(\frac{1+|z|^{-1}}{g^2}+\frac{|z|^{-2}}{g}\right),
 \qquad l\ne k.
\]

To control its diagonal entries, note that $(\partial_zQ)_{kk}$ is
real by $Q_{kk}>0$, whereas
$(Q^{-1}\partial_zQ)_{kk}$ is purely imaginary. Taking imaginary parts in
\[
 (\partial_zQ)_{kk}
 =Q_{kk}(Q^{-1}\partial_zQ)_{kk}
  +\sum_{l\ne k}Q_{kl}(Q^{-1}\partial_zQ)_{lk}
\]
therefore gives
\begin{equation*}
 |(Q^{-1}\partial_zQ)_{kk}|
 =\frac1{Q_{kk}}
 \Big|\operatorname{Im}\sum_{l\ne k}
 Q_{kl}(Q^{-1}\partial_zQ)_{lk}\Big|\le\frac1{Q_{kk}}\sum_{l\ne k}
 |Q_{kl}|\,|(Q^{-1}\partial_zQ)_{lk}|.
\end{equation*}
Since $Q_{kk}\ge 1/2$ for large $t$, the same
bound holds for the diagonal entries. Since $Q$ is unitary for real $z\in I_t$, we get
\begin{equation}\label{Qestimate}
 \|\partial_zQ\|=\|Q^{-1}\partial_zQ\|
 =O\Big(\frac{1+|z|^{-1}}{g^2}+\frac{|z|^{-2}}{g}\Big).
\end{equation}

We then estimate the eigenvalue term. Since $\|N+B/z\|/g=O(L^{-1})$ uniformly on $I_t$,
the second order eigenvalue perturbation formula gives
\begin{equation*}
 \lambda_k=\I u_kz+\I td_k+\frac{B_{kk}}{z}+\sum_{l\ne k}
 \frac{(N_{kl}+B_{kl}/z)(N_{lk}+B_{lk}/z)}
 {\I\bigl((u_k-u_l)z+t(d_k-d_l)\bigr)}
 +O\left(\frac{\|N+B/z\|^3}{g^2}\right).
\end{equation*}
Since $\|N+B/z\|\le C(1+|z|^{-1})$ and
$(1+|z|^{-1})^3=O(1+|z|^{-3})$, the remainder is
$O((1+|z|^{-3})/g^2)$.
Using $N_{kl}N_{lk}=\I(u_k-u_l)q_{kl}$, we obtain
\begin{equation}\label{rhoest}
 \lambda_k=\I u_kz+\I td_k+\frac{B_{kk}}{z}
 +\sum_{l\ne k}\frac{q_{kl}}
 {z+t\frac{d_k-d_l}{u_k-u_l}}+\rho_k,
 \qquad \text{with  }
 |\rho_k|\le C\left(
 \frac{|z|^{-1}+|z|^{-2}}g+\frac{1+|z|^{-3}}{g^2}\right).
\end{equation}
Conservation of $\Lambda_k$ gives
$B_{kk}-\beta_k=-\sum_{l\ne k}(q_{kl}-q^0_{kl})$, so
\begin{equation}\label{lambdaPhiest}
 \lambda_k-\frac{\partial_z\Phi_{kk}}{\Phi_{kk}}
 =\rho_k-\sum_{l\ne k}(q_{kl}-q^0_{kl})
 \Big(\frac1z-\frac1{z+t\frac{d_k-d_l}{u_k-u_l}}\Big).
\end{equation}

To prove \eqref{Qphilambda}, we estimate two integrals, the integral of the absolute value of the difference of fractions in \eqref{lambdaPhiest} and the integral of $\|\partial_zQ\|+\sum_k|\rho_k|$.

For the first integral, choose $R>0$, independent of $t$, sufficiently large that
all crossing points lie in $(-Rt/2,Rt/2)$. On $I_t\cap[-Rt,Rt]$, we use the bounds $|z|\ge L/t$
and $|z-z_a|\ge L$ for every crossing point $z_a$.
For $|z|>Rt$, the difference of the two fractions is
$O(t|z|^{-2})$. Hence we have
\begin{align}\nonumber
 \int_{I_t}
 \Big|\frac{1}{z}-\frac1{z+t\frac{d_k-d_l}{u_k-u_l}}\Big|\,dz&=
 \underbrace{
 \int_{I_t\cap[-Rt,Rt]}
 \Big|\frac{1}{z}-\frac1{z+t\frac{d_k-d_l}{u_k-u_l}}\Big|\,dz
 }_{O(\log t)}\\ \label{zzest}
 &+
 \underbrace{
 \int_{I_t\cap\{|z|>Rt\}}
 \Big|\frac{t\frac{d_k-d_l}{u_k-u_l}}
{z\Big(z+t\frac{d_k-d_l}{u_k-u_l}\Big)}\Big|\,dz
 }_{O(1)}=O(\log t).
\end{align}

For the second integral, the estimates \eqref{Qestimate} and \eqref{rhoest} imply
\[
 \|\partial_z Q\|+\sum_k|\rho_k|
 \le C\Big(
 \frac{|z|^{-1}+|z|^{-2}}{g}+\frac{1+|z|^{-3}}{g^2}
 \Big).
\]
Since zero and the crossing points are separated by distances
of order $t$, we can choose a fixed $c>0$ such that their
neighborhoods of radius $ct$ are disjoint and contained in
$(-Rt,Rt)$. Then we have

\begin{align}\nonumber
\int_{I_t}\Big(\|\partial_z Q\|+\sum_k|\rho_k|\Big)\,dz
 &\le C\underbrace{\int_{L/t\le |z|\le ct}
 \Big(\frac{|z|^{-1}+|z|^{-2}}{t}
       +\frac{1+|z|^{-3}}{t^2}\Big)\,dz}
 _{O(L^{-1}+t^{-1}\log t)}\\ \nonumber
 &+C\underbrace{\sum_{a=1}^m
 \int_{L\le |z-z_a|\le ct}
 \left(\frac{1}{|z-z_a|^2}
       +\frac{1}{t|z-z_a|}\right)\,dz}
 _{O(L^{-1}+t^{-1}\log t)}
\\ \label{Qrhoest}
 &+C\underbrace{\int_{\mathbb R}\frac{dz}{(t+|z|)^2}}_{O(t^{-1})}
 =O(L^{-1}+t^{-1}\log t).
\end{align}
Here the first term bounds the integral over $I_t\cap[-ct,ct]$,
where $g^{-1}=O(t^{-1})$ and $|z|\ge L/t$.
The second term bounds the sum of the integrals over
$I_t\cap[z_a-ct,z_a+ct]$ for $a=1,\ldots,m$, where $g^{-1}=O(|z-z_a|^{-1})$,
$|z|^{-1}=O(t^{-1})$, and $|z-z_a|\ge L$.
The third term bounds the integral over the rest of $I_t$, where the integrand is $O(t^{-2})$ for $|z|\le Rt$ and
$O(|z|^{-2})$ for $|z|>Rt$, and is thus bounded by
$C(t+|z|)^{-2}$.

Since $q_{kl}-q^0_{kl}=O(t^{-1})$,
\eqref{lambdaPhiest}, \eqref{Qrhoest} and \eqref{zzest}
give \eqref{Qphilambda}, and hence \eqref{QPhiest}.

Let $F$ be a solution in the statement on its corresponding interval $I_t$. The coefficient of \eqref{zeq} is skew-Hermitian on $I_t$,
so its propagator is unitary. By the variation of constants formula,
for finite $a,z\in I_t$,
we obtain
\begin{equation*}
 \|F(z,t)-Q(z,t)\Phi(z,t)\|
 \le\|F(a,t)-Q(a,t)\Phi(a,t)\|+\int_{\min \{a,z\}}^{\max\{a,z\}}
 \left\|\partial_s(Q\Phi)-(Us+tD+N+B/s)Q\Phi\right\|\,ds.
\end{equation*}

For a reference solution $\mathcal F_j^\pm(z,t)$, take $a=z_j^\pm$ in this inequality. By the initial condition
$\mathcal F_j^\pm(z_j^\pm,t)=\Phi(z_j^\pm,t)$,
\[
 \mathcal F_j^\pm(z_j^\pm,t)-Q(z_j^\pm,t)\Phi(z_j^\pm,t)
 =(1-Q(z_j^\pm,t))\Phi(z_j^\pm,t)=O(L^{-1}).
\]
For each fixed $t$, \eqref{asymcond} and \eqref{Phiinfinity},
together with $Q(a,t)\to 1$, give
\[
 F_\pm(a,t)-Q(a,t)\Phi(a,t)\longrightarrow 0,
 \qquad a\to \pm\infty.
\]
Therefore, by \eqref{QPhiest}, each solution in
the statement, on its corresponding interval, satisfies
\[
 \sup_{z\in I_t}\|F(z,t)-Q(z,t)\Phi(z,t)\|
 =O(L^{-1}+t^{-1}\log t).
\]

To compare with $\Phi$, we write
\[
 \Phi^{-1}F-1
 =\Phi^{-1}(F-Q\Phi)+\Phi^{-1}(Q-1)\Phi.
\]
Since $Q=1+O(L^{-1})$ uniformly for $z\in I_t$, and
$\Phi,\Phi^{-1}$ are uniformly bounded there, we obtain \eqref{PhiFcomp} and \eqref{PhiEndpointComp}.

As a consequence, for any $z\in[z_j^+,z_{j+1}^-]$, taking the ratio gives
\[
 M_j(t)
 =\bigl(\Phi(z,t)^{-1}\mathcal F_{j+1}^-(z,t)\bigr)^{-1}
   \bigl(\Phi(z,t)^{-1}\mathcal F_j^+(z,t)\bigr)
 =1+O(L^{-1}+t^{-1}\log t).
\]
At $z=z_1^-$, using $\mathcal F_1^-(z_1^-,t)=\Phi(z_1^-,t)$,
we have
\[
 M_0(t)=\Phi(z_1^-,t)^{-1}F_-(z_1^-,t)
 =1+O(L^{-1}+t^{-1}\log t)=1+O(t^{-1/3}).
\]
It finishes the proof.
\end{proof}

\subsection{Matching with the limiting system with a second order pole}
\label{endpointmatching}
It remains to determine $M_+(t)$. 

\begin{pro}\label{proFYcomp}
As $t\to+\infty$,
\begin{align}\label{FYcomp}
 t^{2\beta-\Lambda}\mathcal F_m^+(\zeta/t,t)
 &=Y_-(\zeta)D_{\out}+O(t^{-1/3}), \qquad \zeta<0,\\ \label{FYcomp-}
 t^{2\beta-\Lambda}F_+(\zeta/t,t)
 &=Y_+(\zeta)D_{\out}+O(t^{-1/3}), \qquad \zeta>0.
\end{align}
The estimates are locally uniform for $\zeta<0$ and
$\zeta>0$, respectively. The second
estimate also holds for $\zeta<0$ when both solutions are continued through the upper half plane. Consequently,
\begin{equation}\label{M+limit}
 M_+(t)=D_{\out}^{-1}S_{\out}(D,B_0)D_{\out}+O(t^{-1/3}).
\end{equation}
\end{pro}

\begin{proof}
By \eqref{PhiEndpointComp},
\[
\begin{aligned}
 \Phi(-L/t,t)^{-1}\mathcal F_m^+(-L/t,t)&=1+O(t^{-1/3}),\\
 \Phi(L/t,t)^{-1}F_+(L/t,t)&=1+O(t^{-1/3}).
\end{aligned}
\]
The expansion of \eqref{Phi} at $z=\pm L/t$ gives
\[
 t^{2\beta-\Lambda}\Phi(\pm L/t,t)
 =e^{\pm DL}(\pm L)^{\beta}D_{\out}
 \Big(1+O\Big(\frac{L+L^2}{t^2}\Big)\Big).
\]
Together with the canonical asymptotics of $Y_\pm$,
it implies
\[
\begin{aligned}
 t^{2\beta-\Lambda}\mathcal F_m^+(-L/t,t)
 &=Y_-(-L)D_{\out}+O(t^{-1/3}),\\
 t^{2\beta-\Lambda}F_+(L/t,t)
 &=Y_+(L)D_{\out}+O(t^{-1/3}).
\end{aligned}
\]
Here we use the fact that $e^{\pm DL}(\pm L)^{\beta}D_{\out}$ is bounded
independently of $t$,  and
$t^{2\beta-\Lambda}$ is unitary for $t>0$.

The coefficient matrix of the equation for $t^{2\beta-\Lambda}F(\zeta/t,t)$
differs from $D+B_0/\zeta$ by 
\[
 \frac{U\zeta}{t^2}+\frac{t^{2\beta-\Lambda}N(t)t^{-(2\beta-\Lambda)}}{t}
 +\frac{t^{2\beta-\Lambda}B(t)t^{-(2\beta-\Lambda)}-B_0}{\zeta}.
\]
By boundedness of $N(t)$ and \eqref{Blimit}, the norm of the remainder matrix is at most
$C(|\zeta|/t^2+t^{-1}+(t|\zeta|)^{-1})$.
For fixed $\delta>0$, its integral over either
$[-L,-\delta]$ or $[\delta,L]$ is bounded by
\[
 C\int_\delta^L
 \left(\frac{s}{t^2}+\frac1t+\frac1{ts}\right)\,ds
 =O\Big(\frac{L^2}{t^2}+\frac Lt+\frac{\log L}t\Big)
 =O(t^{-2/3}).
\]
Both equations have skew-Hermitian coefficients on these real
intervals, so their propagators are unitary.
The variation of constants formula proves
\eqref{FYcomp} and
\eqref{FYcomp-}.

In the end, fix $\zeta<0$ and take the continuation along the path $\xi=|\zeta|e^{\I\theta}$, $0\le\theta\le\pi$. The coefficient matrix $D+B_0/\xi$ and the coefficient matrix of the equation for $t^{2\beta-\Lambda}F(\xi/t,t)$ are bounded along the path, and their difference is $O(t^{-1})$. 
Applying the variation of constants formula along the
path, using the estimate at $|\zeta|$, gives
\[
 t^{2\beta-\Lambda}F_+(\zeta/t,t)
 =Y_+(\zeta)D_{\out}+O(t^{-1/3}).
\]
Combining this with \eqref{FYcomp}, we obtain
\[
\begin{aligned}
 M_+(t)
 &=\bigl(t^{2\beta-\Lambda}F_+(\zeta/t,t)\bigr)^{-1}
   t^{2\beta-\Lambda}\mathcal F_m^+(\zeta/t,t)\\
 &=D_{\out}^{-1}Y_+(\zeta)^{-1}Y_-(\zeta)D_{\out}
   +O(t^{-1/3})\\
 &=D_{\out}^{-1}S_{\out}(D,B_0)D_{\out}
   +O(t^{-1/3}).
\end{aligned}
\]
It finishes the proof.
\end{proof}

\subsection{Proof of Theorem \ref{thm2}}
Since $S_0S_1$ is independent of $t$, letting $t\to+\infty$ in \eqref{orderpro}
and using \eqref{Wlimit}, \eqref{Mlimit}, \eqref{M0limit}
and \eqref{M+limit}, we obtain
\begin{equation}
S_0S_1=D_{\out}^{-1}S_{\out}(D,B_0)D_{\out}T^{(p_mq_m)}\cdots T^{(p_1q_1)}.
\end{equation}
It gives a proof of Theorem \ref{thm2}.

\Addresses
\end{document}